\documentclass[aps,prc,twocolumn,10pt, superscriptaddress, showpacs, floatfix]{revtex4-1}

\usepackage{amssymb,epsfig}

\newcommand{\be}{\begin{equation}}
\newcommand{\ee}{\end{equation}}
\newcommand{\bea}{\begin{eqnarray}}
\newcommand{\eea}{\end{eqnarray}}
\begin{document}

\title{Nuclear superfluidity at finite temperature}
\author{Elena Litvinova}
\affiliation{Department of Physics, Western Michigan University, Kalamazoo, MI 49008, USA}
\affiliation{National Superconducting Cyclotron Laboratory, Michigan State University, East Lansing, MI 48824, USA}
\author{Peter Schuck}
\affiliation{Institut de Physique Nucl\'eaire, IN2P3-CNRS, Universit\'e Paris-Sud, F-91406 Orsay Cedex, France}
\affiliation{Universit\'e Grenoble Alpes, CNRS, LPMMC, 38000 Grenoble, France}


\date{\today}

\begin{abstract}
The equation of motion for the two-fermion two-time correlation function in the pairing channel is considered at finite temperature. Within the Matsubara formalism, the Dyson-type Bethe-Salpeter equation (Dyson-BSE) with the frequency-dependent interaction kernel is obtained. Similarly to the case of zero temperature, it is decomposed into the static and dynamical components, where the former is given by the contraction of the bare interaction with the two-fermion density and the latter is represented by the double contraction of the four-fermion two-time correlation function, or propagator, with two interaction matrix elements. The dynamical kernel with the four-body propagator, being formally exact, requires approximations to avoid generating prohibitively complicated hierarchy of equations. 
We focus on the approximation where the dynamical interaction kernel is truncated on the level of two-body correlation functions, neglecting the irreducible three-body and higher-rank correlations. Such a truncation leads to the dynamical kernel with the coupling between correlated fermionic pairs, which can be interpreted as  emergent bosonic quasibound states, or phonons, of normal and superfluid nature. The latter ones are, thus, the mediators of the dynamical superfluid pairing. In this framework, we obtained the closed system of equations for the fermionic particle-hole and particle-particle propagators. This allows us to study the temperature dependence of the pairing gap beyond the Bardeen-Cooper-Schrieffer approximation, that is implemented for medium-heavy nuclear systems. The cases of $^{68}$Ni and $^{44,46}$Ca are discussed in detail.
\end{abstract}

\maketitle

\section{Introduction} 

Superfluidity and superconductivity phenomena in nuclear systems, after their recognition in late 1950's  \cite{Bohr1958}, attracted a tremendous amount of theoretical effort since then. Although the theory of Bardeen, Cooper and Schrieffer (BCS) \cite{Bardeen1957} appeared to be very successful when applied to superconductivity in metals, building a consistent theory of nuclear pairing correlations was complicated by the nature of the nuclear forces.  In particular, the presence of the repulsive core in the nucleon-nucleon interaction and strong in-medium correlations made the direct applicability of the BCS theory to nuclear matter and finite nuclei problematic. The development of powerful many-body methods, such as numerous variants of perturbative and cluster expansions, the correlated basis function method, the Monte Carlo approach and others, together with the advancements of the nucleon-nucleon potentials, has helped significantly to clarify microscopic mechanisms of nuclear superfluidity, eventually going far beyond the BCS theory \cite{50BCS}.

While the observations, such as the rotational anomalies in pulsar periods and measurements of their surface temperatures evidence unambiguously about superfluidity of neutron stars, theoretical models still vary considerably in the description of its characteristics, for instance, the pairing gaps. The use of different nucleon-nucleon interactions and regularization techniques as well as different treatments of polarization effects may cause substantial differences in model predictions.  Refining the models of superfluidity in both symmetric and asymmetric nuclear matters, in particular, clarifying the role of induced pairing in screening and antiscreening is a topic of active research \cite{Shen2005,Cao2006,Ding2016,Ramanan2018,Sedrakian2019,Guo2019a,Urban2020}.

Investigation of pairing correlations in finite nuclei seems to be less intense. In most of the applications to nuclear structure calculations, rather simplistic concepts of pairing like BCS, Hartree-Fock-Bogoliubov or Gor'kov Green functions are employed, which is unavoidable to make otherwise sophisticated many-body calculations feasible. The accuracy of such simplified treatments of pairing is comparable with the errors introduced by other model approximations, such as neglecting high-rank many-body correlations, multiparticle interactions and coupling to the continuum, to name a few.  However, with the progress of those aspects also more accurate treatment of pairing correlations should be considered.

One of the most intriguing issues in strongly-coupled many-body systems is the emergence of collective phenomena. Understanding their significant role in the formation of the nuclear ground and excited states has been progressed impressively over the decades, since Bohr and Mottelson \cite{BohrMottelson1969,BohrMottelson1975}. The impact of collective effects on nuclear pairing was investigated in various phenomenological frameworks \cite{VanderSluys1993,Avdeenkov1999,Avdeenkov1999a,BarrancoBrogliaGoriEtAl1999,BarrancoBortignonBrogliaEtAl2005,IdiniPotelBarrancoEtAl2015}, which revealed that coupling between nucleons and collective surface vibrations (particle-vibration coupling, or PVC) can be responsible for a large fraction of the pairing gap. The PVC effects identified with the major contribution to the induced pairing are widely known to be of prime importance in electronic condensed matter systems, where they induce superconductivity by reversing the sign of the repulsive Coulomb interaction \cite{50BCSorig,50BCS,Strinati2018}.

Atomic nuclei embedded in stellar environments are of special interest. Their response to various changes of those environments and the associated nuclear reactions are complicated by the thermodynamical conditions, first of all, by the non-zero temperature.  It is widely recognized that the modifications of nuclear properties by finite temperature can noticeably influence the star evolution scenarios. In particular, the dependence of nuclear superfluidity on temperature may play a non-negligible role for the electron capture in collapsing stars and r-process nucleosynthesis in neutron star mergers. Whereas the temperature-dependent BCS is well understood and known for the superfluid to non-superfluid phase transition at the critical temperature $T_c \sim 0.6 \Delta(T=0)$, the temperature dependence of the induced pairing is more complicated as well as the induced pairing itself.

This kind of pairing at finite temperature was investigated more intensely for nuclear matter, although abundant shell-model Monte Carlo studies for finite nuclei are available, see Refs. \cite{Langanke1996,LiuAlhassid2001} and references therein.  In recent Ref. \cite{Urban2020} finite-temperature calculations of the singlet pairing gap in dilute neutron matter were performed. The authors investigated the pairing gaps and the critical temperature of the superfluid phase transition. The $V_{\text{low-k}}$ interaction derived from the Argonne potential AV18 was employed for the static kernel of the pairing gap equation and the effective interaction from the Skyrme family was used for the RPA vertices, which determine the dynamical kernel with the induced pairing. At higher densities the full RPA lead to stronger screening than the reference Landau approximation.  As previously, for instance, in the studies of the BEC-BCS crossover and the liquid-gas phase transition in hot and dense nuclear matter \cite{Jin2010}, it was
pointed out that the pairing gap and the phase transition temperature are sensitive to the approximation used to describe the medium polarization effects responsible for the induced pairing.

In the present work we aim at investigating the temperature dependence of the induced pairing in finite nuclei.
Technically, we consider nuclear correlation functions in the equation of motion (EOM) framework, which is one of the most universal methods known across the many areas of quantum physics from condensed matter to  quantum chemistry 
\cite{DukelskyRoepkeSchuck1998,Storozhenko2003,Tiago2008,Martinez2010,Sangalli2011,SchuckTohyama2016,Olevano2018}.
Following our previous developments reported in Ref. \cite{LitvinovaSchuck2020} for the zero-temperature case, in Section \ref{Propagator} we generate the EOM for the fermionic pair, or particle-particle, propagator, but now evolving in the domain of imaginary time introduced by Matsubara \cite{Matsubara1955}.  We show that the four-fermion correlation function in the dynamical kernel of the resulting EOM, which is responsible for the induced pairing, can be approximated with various degrees of accuracy, in analogy with the zero-temperature case. At finite temperature, however, this kernel carries a non-trivial temperature dependence, which is different from that of the static kernel implied in the BCS theory. The impact of the latter temperature dependence on nuclear pairing gaps is studied numerically in Section \ref{Results}. The conclusions are drawn in the summary Section \ref{Conclusions}.

\section{Fermionic pair propagator in a heated correlated medium}
\label{Propagator}
In analogy with Ref. \cite{LitvinovaSchuck2020}, we stay within the formalism of correlation functions, such as the Green functions, or propagators. 
As the propagators are directly related to observed excitation spectra and ground state properties of the many-body systems, this formalism is one of the most convenient and powerful ones in the description of phenomena that occur in strongly-coupled media. Following Matsubara \cite{Matsubara1955}, 
the temperature-dependent propagator of a fermionic pair in a heated correlated medium can be defined as a thermal average \cite{Zagoskin2014}
\bea
&{\cal G}&(12,1'2') \equiv {\cal G}_{12,1'2'}(\tau-\tau') = -\langle T_{\tau} \psi(1)\psi(2){\bar\psi}(2'){\bar\psi}(1')\rangle \nonumber \\
\label{ppGF} 
\eea
with the chronological ordering $T_{\tau}$ of the one-fermion fields in the imaginary time domain of the Wick rotated picture:
\bea
\psi(1) &\equiv& \psi_1(\tau_1) = e^{{\cal H}\tau_1}\psi_1e^{-{\cal H}\tau_1},\nonumber\\
{\bar\psi}(1) &\equiv& {\bar\psi}_1(\tau_1) = e^{{\cal H}\tau_1}{\psi^{\dagger}}_1e^{-{\cal H}\tau_1}.
\label{Wick-Heisenberg}
\eea
The operator ${\cal H}$ is given by ${\cal H} = H - \lambda N$, where $H$ is the many-body Hamiltonian
\be
H = H^{(1)} + V^{(2)} =  \sum_{12} t_{12} \psi^{\dag}_1\psi_2 + \frac{1}{4}\sum\limits_{1234}{\bar v}_{1234}{\psi^{\dagger}}_1{\psi^{\dagger}}_2\psi_4\psi_3,
\label{Hamiltonian}
\ee
with the antisymmetrized matrix elements ${\bar v}_{1234} = v_{1234} - v_{1243}$, $\lambda$ is the chemical potential, and $N$ is the particle number operator.
Here and in the following the number subscript denotes  the full set of the single-particle quantum numbers in a given representation and the imaginary time variables $\tau$ are related to the real times $t$ as  $\tau = it$. The fermionic fields $\psi_1$ and $\psi^{\dagger}_1$ satisfy the usual anticommutation relations, and the angular brackets in Eq. (\ref{ppGF}) stand for the thermal average \cite{Abrikosov1975, Zagoskin2014}
\be
\langle O \rangle = \sum\limits_{\nu} w_{\nu}\langle \nu|O|\nu\rangle
\ee
with the summation over the expectation values in the eigenstates of the Hamiltonian $|\nu\rangle$ weighted with the probabilities $w_{\nu}$ of finding the system in those states within the grand canonical ensemble.
In the present work the Hamiltonian of Eq. (\ref{Hamiltonian}) is confined by the two-body interaction, however, as in the zero-temperature case and as it follows from the discussion below, the generalization to multiparticle forces is straightforward. 


The first equation of motion probing the evolution of the correlation function of a fermionic pair defined by Eq. (\ref{ppGF}) with the imaginary time can be generated by the differentiation of this function with respect to the first time variable $\tau$:
\bea
\partial_{\tau} {\cal G}_{12,1'2'}(\tau-\tau') = -\delta(\tau-\tau')\langle [\psi_1\psi_2,{\psi^{\dagger}}_{2'}{\psi^{\dagger}}_{1'}]\rangle -\nonumber \\
- \langle T_{\tau}[{\cal H},\psi_1\psi_2](\tau)({\bar\psi}_{2'}{\bar\psi}_{1'})(\tau')\rangle, \nonumber\\ 
\label{dtG}                           
\eea
where we adopted the notation:
\be
[{\cal H},A](\tau) = e^{{\cal H}\tau}[{\cal H},A]e^{-{\cal H}\tau}
\ee
for an arbitrary operator $A$. After the evaluation of the commutators
the first EOM reads:
\bea
-(\partial_{\tau} + \varepsilon_1 &+& \varepsilon_2){\cal G}_{12,1'2'}(\tau-\tau') = \delta(\tau-\tau'){\cal N}_{121'2'} + \nonumber \\
&+&\langle  T_{\tau}[V,\psi_1\psi_2](\tau)({\bar\psi}_{2'}{\bar\psi}_{1'})(\tau')\rangle,
\label{EOM1}
\eea
where the single-particle energies $\varepsilon_1$ are $\varepsilon_1 = t_{11} - \lambda$ and we assumed that the working basis diagonalizes the one-body part of the Hamiltonian. The norm matrix in the pp-channel ${\cal N}_{121'2'}$ is the thermal average:
\be
{\cal N}_{121'2'} = \langle [\psi_1\psi_2,{\psi^{\dagger}}_{2'}\psi^{\dagger}_{1'}] \rangle = \delta_{121'2'}(1- n_1 - n_2) = \delta_{121'2'}n_{12},
\label{pp-norm}
\ee
where the one-body density matrix $\rho_{11'}$ obeys $\rho_{11'} = \langle \psi^{\dagger}_{1'} \psi_1 \rangle = \delta_{11'}n_1$ with $n_1$  being, in general, the correlated fermionic occupancies at the given temperature $T$.
In Eq. (\ref{pp-norm}) we adopt the antisymmetrized Kronecker symbol $\delta_{121'2'} = \delta_{11'}\delta_{22'} - \delta_{21'}\delta_{12'}$ and the notation $n_{12} = 1-n_1-n_2$. 

Differentiating the last term on the right hand side of the first EOM (\ref{EOM1}) ${\cal F}_{121'2'}(\tau-\tau') =\langle  T_{\tau}[V,\psi_1\psi_2](\tau)({\bar\psi}_{2'}{\bar\psi}_{1'})(\tau')\rangle$ with respect to the second time argument $\tau'$  generates the second EOM:
\bea
(\partial_{\tau'} - \varepsilon_{1'} - \varepsilon_{2'}){\cal F}_{121'2'}(\tau-\tau') = \nonumber\\ 
= -\delta(\tau-\tau')\langle[[V,\psi_1\psi_2],\psi^{\dagger}_{2'}\psi^{\dagger}_{1'}]\rangle + \nonumber\\ 
+ \langle  T_{\tau}[V,\psi_1\psi_2](\tau)[V,{\bar\psi}_{2'}{\bar\psi}_{1'}](\tau')\rangle .\nonumber\\
\label{EOM2}
\eea
Applying the operator $(\partial_{\tau'} - \varepsilon_{1'} - \varepsilon_{2'})$ to the first EOM (\ref{EOM1}) and combining Eqs. (\ref{EOM1}) and (\ref{EOM2}), we perform the Fourier transformation to the domain of the Matsubara's discrete energy variable $\omega_n = 2\pi nT$. The spectral image in this domain is defined by the relation:
\be
{\cal G}_{12,1'2'}(\tau-\tau') = T\sum\limits_{n}  e^{-i\omega_n(\tau-\tau')}{\cal G}_{12,1'2'}(\omega_n).
\ee 
In this way, we obtain:
\bea
{\cal G}_{12,1'2'}(\omega_n) &=& {\cal G}^{(0)}_{12,1'2'}(\omega_n) + \nonumber \\
&+& \frac{1}{4}\sum\limits_{343'4'}{\cal G}^{(0)}_{12,34}(\omega_n){\cal T}_{343'4'}(\omega_n){\cal G}^{(0)}_{3'4',1'2'}(\omega_n),\nonumber\\  
\label{GTmatrix}
\eea
where the free particle-particle propagator is introduced as:
\be
{\cal G}^{(0)}_{12,1'2'}(\omega_n) = \frac{{\cal N}_{121'2'}}{i\omega_n - \varepsilon_1 - \varepsilon_2}.
\label{ppuncor}
\ee
The interaction  kernel of Eq. (\ref{GTmatrix}) has the meaning of $T$-matrix and reads:
\be
{\cal T}_{121'2'}(\omega_n) =  \frac{1}{4}\sum\limits_{343'4'}{\cal N}^{-1}_{1234}\Bigl( {\cal T}^{(0)}_{343'4'} + {\cal T}^{(r)}_{343'4'}(\omega_n)\Bigr){\cal N}^{-1}_{3'4'1'2'}.
\ee
The components ${\cal T}^{(0)}_{343'4'} $ and ${\cal T}^{(r)}_{343'4'}(\omega_n)$ are, formally,  the Fourier images of the two last terms on the right hand side of Eq. (\ref{EOM2}), i.e.,
\bea
{\cal T}^{(0)}_{121'2'}(\tau-\tau') &=&  -\delta(\tau-\tau') \langle  [[V,\psi_1\psi_2],{\psi^{\dagger}}_{2'}{\psi^{\dagger}}_{1'}]\rangle\nonumber\\
{\cal T}^{(r)}_{121'2'}(\tau-\tau') &=&  \langle  T_{\tau}[V,\psi_1\psi_2](\tau)[V,{{\bar\psi}}_{2'}{{\bar\psi}}_{1'}](\tau')\rangle, \nonumber\\
\label{Tstdyn}
\eea
so that ${\cal T}^{(0)}$ is the instantaneous, or static, part of the $\cal T$-matrix and ${\cal T}^{(r)}$ is its dynamical part.
In analogy with the case of the particle-hole response \cite{LitvinovaSchuck2019} and the zero-temperature particle-particle response \cite{LitvinovaSchuck2020}, Eq. (\ref{GTmatrix}) can be transformed to an equation of the Dyson type
\bea
{\cal G}_{12,1'2'}(\omega_n) &=& {\cal G}^{(0)}_{12,1'2'}(\omega_n) + \nonumber \\
&+& \frac{1}{4}\sum\limits_{343'4'}{\cal G}^{(0)}_{12,34}(\omega_n){\cal K}_{343'4'}(\omega_n){\cal G}_{3'4',1'2'}(\omega_n)\nonumber\\  
\label{GDyson}
\eea
by introducing the kernel ${\cal K}(\omega)$ irreducible with respect to the uncorrelated pp-propagator 
(\ref{ppuncor}): 
\bea
{\cal T}_{121'2'}(\omega_n) &=& {\cal K}_{121'2'}(\omega_n) + \nonumber \\
&+& \frac{1}{4}\sum\limits_{343'4'}{\cal K}_{1234}(\omega_n){\cal G}^{(0)}_{34,3'4'}(\omega_n){\cal T}_{3'4'1'2'}(\omega_n),\nonumber\\  
\label{Tmatrix}
\eea
or ${\cal K}(\omega_n) = {\cal T}^{(irr)}(\omega_n)$. 
Thus, at finite temperature the EOM for the propagator of a fermionic pair also acquires the form of the Dyson Bethe-Salpeter equation (Dyson-BSE) \cite{Schuck2019}. To further specify the interaction kernel of the latter equation, one has to evaluate the commutators of Eq. (\ref{Tstdyn}). For the static part, we find, similarly to the case of $T = 0$   \cite{DukelskyRoepkeSchuck1998,LitvinovaSchuck2020}:
\bea 
{\cal T}^{(0)}_{121'2'} &=& \delta_{121'2'}n_{12}({\tilde\Sigma}_1 + {\tilde\Sigma}_2) + {\cal K}^{(0)}_{121'2'}, \label{T02}\\
{\cal K}^{(0)}_{121'2'} &=& {\bar v}_{121'2'}n_{12}n_{1'2'}  \nonumber \\
&-& \Bigl[ \Bigl( \sum\limits_{il}{\bar v}_{i12'l}\sigma^{(2)}_{l2i1'} + \frac{\delta_{22'}}{2}\sum\limits_{ikl} {\bar v}_{i1kl}\sigma^{(2)}_{kli1'}\Bigr)  \nonumber\\
&-& \Bigl(1'\leftrightarrow 2' \Bigr) \Bigr] -\Bigl[1\leftrightarrow 2 \Bigr], 
\label{K0}
\eea
where $\sigma^{(2)}_{ijkl}$ is the correlated part of the two-body density
\be
\rho_{ijkl} = \langle \psi^{\dagger}_{k}\psi^{\dagger}_{l}\psi_{j}\psi_{i}\rangle = \rho_{ik}\rho_{jl} - \rho_{il}\rho_{jk} + \sigma^{(2)}_{ijkl}
\ee
and ${\tilde\Sigma}_{11'}$ is the mean-field part of the single-particle self-energy 
\be
{\tilde\Sigma}_{11'} = \sum\limits_{l}{\bar v}_{1l1'l}n_l, \ \ \ \ \ \ \  {\tilde\Sigma}_{11'} = {\delta}_{11'}{\tilde\Sigma}_{1}.
\ee
The latter ones can be absorbed in the uncorrelated propagator, so that the Dyson-BSE takes the form:
\bea
{\cal G}_{12,1'2'}(\omega_n) &=& {\tilde{\cal G}}^{(0)}_{12,1'2'}(\omega_n) + \nonumber \\
&+& \frac{1}{4}\sum\limits_{343'4'}{\tilde{\cal G}}^{(0)}_{12,34}(\omega_n){\cal K}_{343'4'}(\omega_n){\cal G}_{3'4',1'2'}(\omega_n),\nonumber\\  
\label{GDyson1}
\eea
where ${\tilde{\cal G}}^{(0)}_{12,1'2'}(\omega_n)$ is the uncorrelated particle-particle propagator in the mean field
\be
{\tilde{\cal G}}^{(0)}_{12,1'2'}(\omega_n) = \frac{\tilde{\cal N}_{121'2'}}{i\omega_n - {\tilde\varepsilon}_1 - {\tilde\varepsilon}_2}, \ \ \ \ \ \ {\tilde\varepsilon}_1 = {\varepsilon}_1 + {\tilde\Sigma}_{1},
\label{ppuncortilde}
\ee
while the kernel $\cal K$ does not contain the mean-field term in its static part. The full interaction kernel of Eq. (\ref{GDyson1}) can then be written as: ${\cal K} = \tilde{\cal N}^{-1}({\cal K}^{(0)} + {\cal K}^{(r)})\tilde{\cal N}^{-1}$, and the symbol "$ \tilde{\ \ \ } $" indicates the mean-field character of the quantity. Remarkably, the static kernel has the same form as at $T =0$ because of its instantaneous character, however, it depends implicitly on temperature via the fermionic densities. 

The  presence of the static kernel is a direct consequence of the instantaneous nature of the bare interaction $\bar v$, that was our initial assumption about the Hamiltonian (\ref{Hamiltonian}). In general, the fermionic bare interaction does not have to be instantaneous, for instance, it can be mediated by a boson, whose exchange between fermions must have retardation. In the case of nuclear forces, that is the meson exchange between two nucleons in the vacuum. We will see in the following that the in-medium analog of this type of interaction can be generated in the present EOM framework, if the dynamical kernel is taken into account. Certain approximations, such as cluster decompositions of the dynamical kernel, bring its structure to the boson-exchange form, where the emergent bosons are correlated fermionic pairs, and the intermediate propagators of these bosons are associated with retardation effects of the in-medium induced interation. As we showed in detail in Refs. \cite{LitvinovaSchuck2019,LitvinovaSchuck2020}, the in-medium dynamical kernel in the form of the phonon-exchange interaction is completely analogous to the meson-exchange interaction between the nucleons in the vacuum. 
In turn, the latter interaction should be, in principle, derivable from the Hamiltonian of quarks and gluons. However, if one wants to rely on the scale separation and consider nucleons as elementary degrees of freedom, the consistent framework implies neglecting the time-dependence of the meson-exchange interaction. The latter approximation is widely used in the low-energy nuclear physics in the so-called "ab initio" calculations.  

The most common practice  for various applications of the many-body theory is to treat the EOM (\ref{GDyson1}) in the simplest approximation, which retains only the static part ${\cal K}^{(0)}$ of the kernel and neglects the correlations originating from ${\cal K}^{(r)}$. Such an approach forms the content of the self-consistent particle-particle random phase approximation (RPA) \cite{RingSchuck1980}, the analogous particle-hole RPA and the self-consistent quasiparticle random phase approximation, or SCQRPA, which combines both of them and demonstrates great success in applications to exactly-solvable models \cite{Rabhi2002}. In nuclear physics, moreover, the common practice is not to compute the static kernel according to Eq. (\ref{K0}), but rather use effective interactions adjusted to finite nuclei or the G-matrix of the Br\"uckner's type.

More and more applications of such approaches to nuclear systems, however, indicate that confining by only the static part of the interaction kernel can not lead to satisfactory results. The most spectacular examples are nuclear excitation spectra and associated decay properties, where the part ${\cal K}^{(r)}$ of the kernel associated with dynamical processes induced by the medium plays a decisive role \cite{LitvinovaSchuck2019,RobinLitvinova2016,Robin2019}.
In the description of superfluid nuclear matter, this part of the kernel produces an interplay of screening and antiscreening effects which can be revealed, for instance, in the calculations of pairing gaps \cite{Cao2006,Guo2019a}. 

The evaluation of the commutators determining the dynamical kernel leads to the following result:
\bea
&\ &{\cal K}^{(r)}_{121'2'}(\tau-\tau') = \frac{1}{4}\times\nonumber\\
&\times&\sum\limits_{ikl}\sum\limits_{mnq}\Bigl[{\bar v}_{i1kl}\langle T_{\tau}(\psi^{\dagger}_i\psi_2\psi_l\psi_k)(\tau)(\psi^{\dagger}_m\psi^{\dagger}_n\psi^{\dagger}_{2'}\psi_q)(\tau')\rangle^{irr} {\bar v}_{mn1'q}   \nonumber\\
&\ &+ {\bar v}_{i1kl}\langle T_{\tau}(\psi^{\dagger}_i\psi_2\psi_l\psi_k)(\tau)(\psi^{\dagger}_m\psi^{\dagger}_n\psi_q\psi^{\dagger}_{1'})(\tau')\rangle^{irr} {\bar v}_{mn2'q}
 \nonumber \\
&\ &+ {\bar v}_{i2kl}\langle T_{\tau}(\psi_1\psi^{\dagger}_i\psi_l\psi_k)(\tau)(\psi^{\dagger}_m\psi^{\dagger}_n\psi^{\dagger}_{2'}\psi_q)(\tau')\rangle^{irr} {\bar v}_{mn1'q} \nonumber\\
&\ &+ {\bar v}_{i2kl}\langle T_{\tau}(\psi_1\psi^{\dagger}_i\psi_l\psi_k)(\tau)(\psi^{\dagger}_m\psi^{\dagger}_n\psi_q\psi^{\dagger}_{1'})(\tau')\rangle^{irr} {\bar v}_{mn2'q}\Bigr]  \nonumber\\
&=& {\cal K}^{(r;11)}_{121'2'}(\tau-\tau') + {\cal K}^{(r;12)}_{121'2'}(\tau-\tau')  \nonumber \\
&+ &{\cal K}^{(r;21)}_{121'2'}(\tau-\tau') + {\cal K}^{(r;22)}_{121'2'}(\tau-\tau')
\label{Tr}
\eea
which, in complete analogy to the case of the particle-particle propagator at zero temperature \cite{LitvinovaSchuck2020}, is determined by the 
two-time four-fermion correlation functions contracted with two interaction matrix elements  in all possible ways, which lead to a four-leg interaction kernel. Each term of Eq. (\ref{Tr}) contains a propagator of three particles and one hole ($3p-1h$). Rather than generating new EOM's for such higher-rank propagators, we will follow the approach of cluster decomposition of Eq. (\ref{Tr}) including up to two-fermion correlation functions:
\bea
{\cal K}^{(r;11)}_{121'2'}(\tau-\tau') = \frac{1}{4}\sum\limits_{ikl}\sum\limits_{mnq}{\bar v}_{i1kl}\times\nonumber\\
\times \Bigl([{\cal R}_{i2,q2'}{\cal G}_{lk,nm}](\tau-\tau') + [{\cal R}_{ik,qn}{\cal G}_{l2,2'm}](\tau-\tau') + \nonumber \\ + [{\cal R}_{il,qm}{\cal G}_{k2,2'n}](\tau-\tau') 
-{\cal AS}\Bigr) {\bar v}_{mn1'q},\nonumber\\
\label{K11}
\eea
\bea
{\cal K}^{(r;12)}_{121'2'}(\tau-\tau') = -\frac{1}{4}\sum\limits_{ikl}\sum\limits_{mnq}{\bar v}_{i1kl}\times\nonumber\\
\times\Bigl([{\cal R}_{i2,qn}{\cal G}_{lk,m1'}](\tau-\tau') + [{\cal R}_{il,q1'}{\cal G}_{k2,nm}](\tau-\tau') +  \nonumber \\ + [{\cal R}_{ik,qm}{\cal G}_{l2,n1'}](\tau-\tau') 
-{\cal AS}\Bigr) {\bar v}_{mn2'q},\nonumber\\
\label{K12}
\eea
\be
{\cal K}^{(r;21)}_{121'2'}(\tau-\tau') = {\cal K}^{(r;12)}_{212'1'}(\tau-\tau'),
\label{K21}
\ee
\be
{\cal K}^{(r;22)}_{121'2'}(\tau-\tau') = {\cal K}^{(r;11)}_{212'1'}(\tau-\tau'), 
\label{K22}
\ee
where we implied that $[{\cal R}_{i2,q2'}{\cal G}_{lk,nm}](\tau-\tau') \equiv {\cal R}_{i2,q2'}(\tau-\tau'){\cal G}_{lk,nm}(\tau-\tau')$ and the finite-temperature particle-hole response function $\cal R$ is introduced as
\bea
&{\cal R}&(12,1'2') \equiv {\cal R}_{12,1'2'}(\tau-\tau') = -\langle T_{\tau} {\bar\psi}(1)\psi(2){\bar\psi}(2'){\psi}(1')\rangle \nonumber.\\
\label{phGF} 
\eea
Eqs. (\ref{K11} - \ref{K22}) are, again, completely analogous to the zero-temperature case \cite{LitvinovaSchuck2020}. The Fourier transformation of the latter kernel to the domain of the Matsubara frequencies requires calculation of the following generic integral
\be
[{\cal R}_{12,1'2'}{\cal G}_{34,3'4'}](\omega_n) = \int\limits_{-1/T}^{1/T}d\tau e^{i\omega_n\tau} {\cal R}_{12,1'2'}(\tau){\cal G}_{34,3'4'}(\tau),  
\label{RGFtransform}
\ee
which yields:
\bea
[{\cal R}_{12,1'2'}{\cal G}_{34,3'4'}](\omega_n) = \nonumber\\
= \sum\limits_{\nu'\nu^{\prime\prime}} w_{\nu\prime}w_{\nu^{\prime\prime}}\Bigl[\sum\limits_{\nu\mu}\frac{ \rho_{21}^{\nu\nu'}\rho_{2'1'}^{\nu\nu'\ast}\alpha_{43}^{\mu\nu^{\prime\prime}}\alpha_{4'3'}^{\mu\nu^{\prime\prime}\ast} }{i\omega_n - \omega_{\nu\nu'} - \omega_{\mu\nu^{\prime\prime}}^{(++)} }\bigl(e^{-(\omega_{\nu\nu'} + \omega_{\mu\nu^{\prime\prime}}^{(++)})/T} - 1\bigr) \nonumber\\
- \sum\limits_{\nu\varkappa}\frac{ \rho_{12}^{\nu\nu'\ast}\rho_{1'2'}^{\nu\nu'}\beta_{34}^{\varkappa\nu^{\prime\prime}\ast}\beta_{3'4'}^{\varkappa\nu^{\prime\prime}} }{i\omega_n + \omega_{\nu\nu'} + \omega_{\varkappa\nu^{\prime\prime}}^{(--)}}\bigl(e^{-(\omega_{\nu\nu'} + \omega_{\varkappa\nu^{\prime\prime}}^{(--)})/T} - 1\bigr)\Bigr].\nonumber\\
\label{RG}
\eea   
In Eq. (\ref{RG}) we have introduced the matrix elements of the normal $ \rho_{12}^{\nu\nu'}$ and pairing $ \alpha_{12}^{\mu\nu}, \beta_{12}^{\varkappa\nu}$ transition densities:
\bea
 \rho_{12}^{\nu\nu'} &=& \langle \nu'|\psi^{\dagger}_2\psi_1|\nu \rangle \nonumber\\
 \alpha_{12}^{\mu\nu} = \langle \nu^{(N)} | \psi_2\psi_1|\mu^{(N+2)} \rangle , \ \ &\ &\ \  \beta_{12}^{\varkappa\nu} = \langle \nu^{(N)} | \psi^{\dagger}_2\psi^{\dagger}_1|\varkappa^{(N-2)} \rangle , \nonumber\\
\eea
where the former connects the states $|\nu\rangle$ of the given $N$-particle system and the latter connect the states  $|\nu\rangle$ with the states $|\mu\rangle , |\varkappa\rangle$ of the $N\pm 2$-particle systems, respectively. The frequencies in the denominators correspond to the associated energy differences.
Similarly to Refs. \cite{LitvinovaSchuck2019,LitvinovaSchuck2020}, it is convenient to also introduce the vertices of the emergent normal and pairing phonons as follows:
\bea
g^{\nu\nu'}_{13} &=& \sum\limits_{24}{\bar v}_{1234}\rho^{\nu\nu'}_{42}, \nonumber\\
\gamma^{\mu\nu(+)}_{12} = \sum\limits_{34} v_{1234}\alpha_{34}^{\mu\nu}, \ \ \ &\ &\ \ \gamma_{12}^{\varkappa\nu(-)} = \sum\limits_{34}\beta_{34}^{\varkappa\nu}v_{3412}, \nonumber\\
\label{phonon}
\eea 
where the presence of the two upper indices indicates that these vertices characterize transitions between excited states, in contrast to the case of zero temperature, where only transitions between the ground and excited states were considered. 
Then, the first component of the dynamical kernel takes the following form:  
\bea
{\cal K}^{(r;11)}_{121'2'}(\omega_n) = -\sum\limits_{\nu^{\prime}\nu^{\prime\prime}} w_{\nu\prime}w_{\nu^{\prime\prime}}\nonumber \\
\times\Bigl[ 
\sum\limits_{\nu\mu}\frac{\Theta^{\mu\nu;\nu^{\prime}\nu^{\prime\prime}(+)}_{121'2'}}{i\omega_n - \omega_{\nu\nu'} - \omega_{\mu\nu^{\prime\prime}}^{(++)} }
\bigl(e^{-(\omega_{\nu\nu'} + \omega_{\mu\nu^{\prime\prime}}^{(++)})/T} - 1\bigr) \nonumber\\ 
-\sum\limits_{\nu\varkappa}\frac{\Theta^{\varkappa\nu;\nu^{\prime}\nu^{\prime\prime}(-)}_{121'2'}}{i\omega_n + \omega_{\nu\nu'} + \omega_{\varkappa\nu^{\prime\prime}}^{(--)} }
\bigl(e^{-(\omega_{\nu\nu'} + \omega_{\varkappa\nu^{\prime\prime}}^{(--)})/T} - 1\bigr)
 \Bigr]
\nonumber\\
\label{K11phon}
\eea
with
\bea
\Theta^{\mu\nu;\nu^{\prime}\nu^{\prime\prime}(+)}_{121'2'} &=& \nonumber\\
= \sum\limits_{kn} \bigl( g^{\nu\nu'}_{1k}\alpha^{\mu\nu^{\prime\prime}}_{2k}\alpha^{\mu\nu^{\prime\prime}\ast}_{2'n}g^{\nu\nu'\ast}_{1'n} &+& \gamma^{\mu\nu^{\prime\prime(+)}}_{1k}\rho^{\nu\nu'}_{2k}\rho^{\nu\nu'\ast}_{2'n}\gamma^{\mu\nu^{\prime\prime(+)}\ast}_{1'n}\bigr) \nonumber\\
\Theta^{\varkappa\nu;\nu^{\prime}\nu^{\prime\prime}(-)}_{121'2'} &=& \nonumber\\
= \sum\limits_{kn} \bigl( g^{\nu\nu'\ast}_{k1}\beta^{\varkappa\nu^{\prime\prime}\ast}_{k2}\beta^{\varkappa\nu^{\prime\prime}}_{n2'}g^{\nu\nu'}_{n1'} &+& \gamma^{\varkappa\nu^{\prime\prime(-)}\ast}_{k1}\rho^{\nu\nu'\ast}_{k2}\rho^{\nu\nu'}_{n2'}\gamma^{\varkappa\nu^{\prime\prime(-)}}_{n1'}\bigr), \nonumber\\
\label{resK11}
\eea
while the second component reads:
\bea
{\cal K}^{(r;12)}_{121'2'}(\omega_n) = \sum\limits_{\nu^{\prime}\nu^{\prime\prime}} w_{\nu\prime}w_{\nu^{\prime\prime}}\nonumber \\
\times\Bigl[ 
\sum\limits_{\nu\mu}\frac{\Sigma^{\mu\nu;\nu^{\prime}\nu^{\prime\prime}(+)}_{121'2'}}{i\omega_n - \omega_{\nu\nu'} - \omega_{\mu\nu^{\prime\prime}}^{(++)} }
\bigl(e^{-(\omega_{\nu\nu'} + \omega_{\mu\nu^{\prime\prime}}^{(++)})/T} - 1\bigr) \nonumber\\ 
-\sum\limits_{\nu\varkappa}\frac{\Sigma^{\varkappa\nu;\nu^{\prime}\nu^{\prime\prime}(-)}_{121'2'}}{i\omega_n + \omega_{\nu\nu'} + \omega_{\varkappa\nu^{\prime\prime}}^{(--)} }
\bigl(e^{-(\omega_{\nu\nu'} + \omega_{\varkappa\nu^{\prime\prime}}^{(--)})/T} - 1\bigr)
 \Bigr],
\nonumber\\
\label{K12phon}
\eea
where
\bea
\Sigma^{\mu\nu;\nu^{\prime}\nu^{\prime\prime}(+)}_{121'2'} = \sum\limits_{ik} \bigl(
\gamma_{1i}^{\mu\nu^{\prime\prime}(+)}\rho_{2i}^{\nu\nu'}\alpha_{1'k}^{\mu\nu^{\prime\prime}\ast} g_{2'k}^{\nu\nu'\ast}  \nonumber \\ 
+ g_{1i}^{\nu\nu'}\alpha_{2i}^{\mu\nu^{\prime\prime}} \rho_{1'k}^{\nu\nu'\ast}\gamma_{2'k}^{\mu\nu^{\prime\prime}(+)\ast}
+ g_{1i}^{\nu\nu'}\alpha_{2i}^{\mu\nu^{\prime\prime}}\alpha_{1'k}^{\mu\nu^{\prime\prime}\ast} g_{2'k}^{\nu\nu'\ast} \bigr)\nonumber \\
\Sigma^{\varkappa\nu;\nu^{\prime}\nu^{\prime\prime}(-)}_{121'2'} = \sum\limits_{ik} \bigl(
\gamma_{i1}^{\varkappa\nu^{\prime\prime}(-)\ast}\rho_{i2}^{\nu\nu'\ast}\beta_{k1'}^{\varkappa\nu^{\prime\prime}} g_{k2'}^{\nu\nu'}  \nonumber \\ 
+ g_{i1}^{\nu\nu'\ast}\beta_{i2}^{\varkappa\nu^{\prime\prime}\ast} \rho_{k1'}^{\nu\nu'}\gamma_{k2'}^{\varkappa\nu^{\prime\prime}(-)}
+ g_{i1}^{\nu\nu'\ast}\beta_{i2}^{\varkappa\nu^{\prime\prime}\ast}\beta_{k1'}^{\varkappa\nu^{\prime\prime}} g_{k2'}^{\nu\nu'} \bigr).\nonumber \\
\label{resK12}
\eea
 The two remaining components ${\cal K}^{(r;21)}_{121'2'}(\omega_n)$ and ${\cal K}^{(r;22)}_{121'2'}(\omega_n)$ can be found from Eqs. (\ref{K11phon} - \ref{resK12}) with the help of the symmetry relations of Eqs. (\ref{K21},\ref{K22}).  Finally, all the expressions can be analytically continued to the domain of real energies.

It is easy to see that in the approximation of Eq. (\ref{K11} - \ref{K22}) to the dynamical kernel the many-body problem can be formulated as a closed scheme. For that, one would need to supplement Eq. (\ref{GDyson1}) with an analogous EOM for the particle-hole response function (\ref{phGF}) and for the lower-rank single-fermion propagator. With the cluster decomposition confined by the two-fermion propagators, all these propagators can be, in principle, found by a self-consistent iterative procedure. Possibilities to implement such a program for nuclear systems will be investigated elsewhere, and for the rest of this work we will focus on calculations of the quantity which is commonly referred to as pairing gap. 

The equation for the pairing gap can be obtained from Eq. (\ref{GDyson1}) if, for instance, the energy argument of the pair propagator is close to the transition frequency from the ground state of the $N$-particle system to the ground state of the $(N+2)$-particle system. Then, the equation for the pairing transition density  $\alpha^{\mu\nu} = \alpha^{s}$ in the vicinity of this frequency $\omega_s$ reads:
\be
\alpha^s_{21} = \frac{1-n_1 - n_2}{\omega_s - {\tilde\varepsilon}_1 - {\tilde\varepsilon}_2} \frac{1}{4}\sum\limits_{343'4'}\delta_{1234}{\cal K}_{343'4'}(\omega_s)\alpha^s_{4'3'}.
\label{alpha}
\ee
If we assume that the ground state of the reference nucleus is approximated by the finite-temperature BCS-like variational ansatz, where
\bea
n_1(T) = v_1^2(1 - f_1(T)) + (1 - v_1^2)f_1(T), \\ 
f_1(T) = \frac{1}{\text{exp}(E_1/T) + 1}, \\ 
E_1 = \sqrt{({\tilde\varepsilon}_1-{\tilde\lambda})^2+\Delta_1^2 } \ \ \ \ \ \ \ v_1^2 = \frac{E_1-({\tilde\varepsilon}_1-{\tilde\lambda})}{2E_1}, \nonumber\\ 
\eea
the finite-temperature $S$-wave pairing gap $\Delta_1$ can be related to the pairing transition density as 
\be
\Delta_1(T) = \alpha^{s}_{{\bar 1}1}\frac{2E_1}{1 - 2f_1(T)}, 
\ee
where the bar denotes the conjugate or the time-reversed state \cite{RingSchuck1980}.
In this approximation, at the frequency $\omega_s = 2{\tilde\lambda}$ Eq. (\ref{alpha}) takes the form of the well-known pairing gap equation:
\be
\Delta_1(T) = -\sum\limits_{2}{\cal V}_{1{\bar 1}2{\bar 2}}\frac{\Delta_2(T)(1-2f_2(T))}{2E_2},
\label{gap}
\ee
which has formally the same structure as the finite-temperature BCS equation, but with a more complicated interaction kernel
\be
{\cal V}_{121'2'} =  \frac{1}{2}\Bigl({\cal K}^{(0)}_{121'2'} + {\cal K}^{(r)}_{121'2'}(2{\tilde\lambda}) \Bigr),
\label{Kgap}
\ee 
whose both the static and dynamical components include the $\langle {\tilde{\cal N}}^{-1}...\ {\tilde{\cal N}}^{-1}\rangle$ factors. The dynamical part, although taken in the static limit, carries the retardation effects and the additional temperature dependence. Notice that Eq. (\ref{gap}) has the same form regardless the approximations made for its static ${\cal K}^{(0)}$ and dynamical ${\cal K}^{(r)}$ parts. As the fully self-consistent treatment of those kernels is difficult even in the approximation made for ${\cal K}^{(r)}$ in Eqs. (\ref{K11} - \ref{K22}), further approximations can be made. Besides the most strong BCS-like one neglecting correlations, such as the complete dynamical part  ${\cal K}^{(r)}$ and the terms with $\sigma^{(2)}$ in ${\cal K}^{(0)}$, one can make weaker approximations. For the static kernel ${\cal K}^{(0)}$ this could be the G-matrix, various kinds of preprocessing of the bare interactions, such as the renormalization group or low-k, and, eventually, effective interactions. For the dynamical kernel ${\cal K}^{(r)}$ the ${\cal R}$, ${\cal G}$ or both correlation functions appearing in Eqs. (\ref{K11} - \ref{K22}) can be approximated by their uncorrelated mean-field analogs or, alternatively, correlations in these propagators can be only partly relaxed. This type of approaches were applied, for instance, for the nuclear matter calculations of Refs. \cite{Ainsworth1989,Schulze1996,Cao2006}, to name a few. 

\section{Details of calculations, results and discussion}
\label{Results}

The numerical implementation of the approach of Eq. (\ref{gap}) for the pairing gap with the kernel of Eq. (\ref{Kgap}) aimed at the investigation of the temperature dependence of the induced pairing, i.e., essentially of the role of the second term in Eq. (\ref{Kgap}). Therefore, at this point we kept the static kernel as simple as in Ref. \cite{LitvinovaSchuck2020}, namely described by the effective monopole-monopole force with adjustable strength to avoid complications like taming the bare interaction with the hard core. The latter will be investigated elsewhere. 

As in the previous implementations of the relativistic finite-temperature approaches with PVC, first we solve the closed set of the relativistic mean field (RMF) equations using the non-linear sigma-model and the NL3 parametrization \cite{Lalazissis1997}, where the Fermi-Dirac thermal fermionic occupation numbers self-consistently modify the classical meson fields. The procedure generates a set of temperature-dependent single-particle Dirac spinors and the corresponding single-nucleon energies, which serve as the working basis $\{1,{\tilde\varepsilon}_1\}$. 
Then, the finite-temperature relativistic random phase approximation (FT-RRPA) equations are solved to obtain the phonon vertices $g^{m}\equiv g^{\nu\nu'}$ and their frequencies $\omega_{m}\equiv \omega_{\nu\nu'}$. In this implementation we relaxed correlations in the particle-particle propagator of Eqs. (\ref{K11}-\ref{K22}), because they are known to be less important than the correlations in the particle-hole propagator. This means, technically, (i) neglecting the terms with $\gamma$-vertices in the Eqs. (\ref{resK11},\ref{resK12}) and (ii) replacing the pairing transition densities $\alpha$ and $\beta$ with their uncorrelated analogs and, simultaneously, the pairing frequencies $\omega^{(\pm\pm)}$ with the sums of the single-particle energies. Thus, the model space for the dynamical kernel is formed by
the set of the obtained FT-RRPA phonons and the thermal RMF single-particle states 
coupled in the $pp\otimes$phonon, $hh\otimes$phonon and $ph\otimes$phonon configurations. To avoid the divergencies of the norm factors $\langle {\tilde{\cal N}}^{-1}...\ {\tilde{\cal N}}^{-1}\rangle$ around the Fermi energy, their $T = 0$ Hartree values were used in calculations. This approximation may be relaxed in the self-consistent calculations of the dynamical kernel, if pairing correlations are included explicitly in ${\cal K}^{(r)}$. 

Particle-hole configurations with the energies $\varepsilon_{ph}\leq100$ MeV and the antiparticle-hole ($\alpha h$) ones with $\varepsilon_{\alpha h}\geq-1800$ MeV, with respect to the positive-energy continuum, limited the particle-hole basis for the FT-RRPA calculations of the phonons. The set of phonons included vibrations with spins and parities $J^{\pi}=2^{+},\;3^{-},\;4^{+},\;5^{-},\;6^{+}$ below 20 MeV. The phonon modes with the reduced transition probabilities $B(EL)$ equal or more than 5\% of the maximal one (for each $J^{\pi}$) were included in the model space, and
the single-particle intermediate states entering the matrix elements $g^{\nu\nu'}_{nk}$ in the summations of Eqs. (\ref{resK11},\ref{resK12}) with the energy differences  $|\varepsilon_{k}- \varepsilon_{n}| \leq$ 50 MeV were included in the summation. 
The same truncation criteria were applied for all temperature regimes, that is justified by our previous calculations \cite{LitvinovaWibowo2018,WibowoLitvinova2019,LitvinovaWibowo2019}. Note that we did not take into account the phenomenological static pairing in the mean field calculations and in the calculations of the PVC vertices as it was done in the first application of the approach at $T = 0$ in Ref. \cite{LitvinovaSchuck2020}, therefore, the results at zero temperature are slightly different (see a more detailed discussion below). The main features of the solutions, however, remain intact.
\begin{figure}
\begin{center}
\includegraphics[scale=0.33]{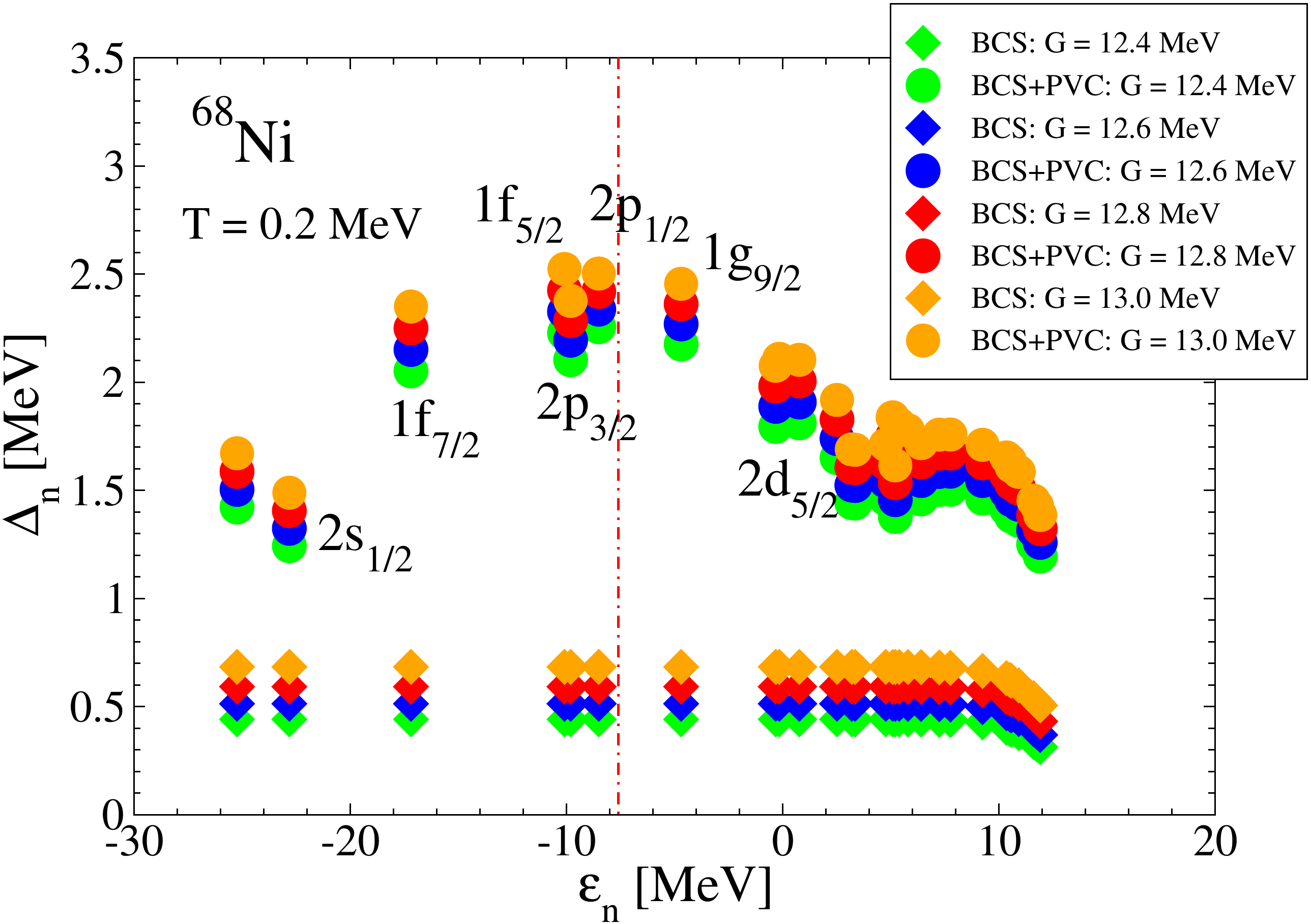}
\end{center}
\caption{Sensitivity of the neutron pairing gap around the Fermi surface of $^{68}$Ni to the strength of the static interaction: BCS and BCS+PVC calculations with varying static pairing strength are shown by diamonds and circles, respectively.  In both approaches, larger values correspond to larger static pairing  strength, see text for details. The vertical line marks the Fermi energy in the BCS+PVC calculations at $T = 0$.}
\label{g-sensitivity}%
\end{figure}

\begin{figure}
\begin{center}
\includegraphics[scale=0.37]{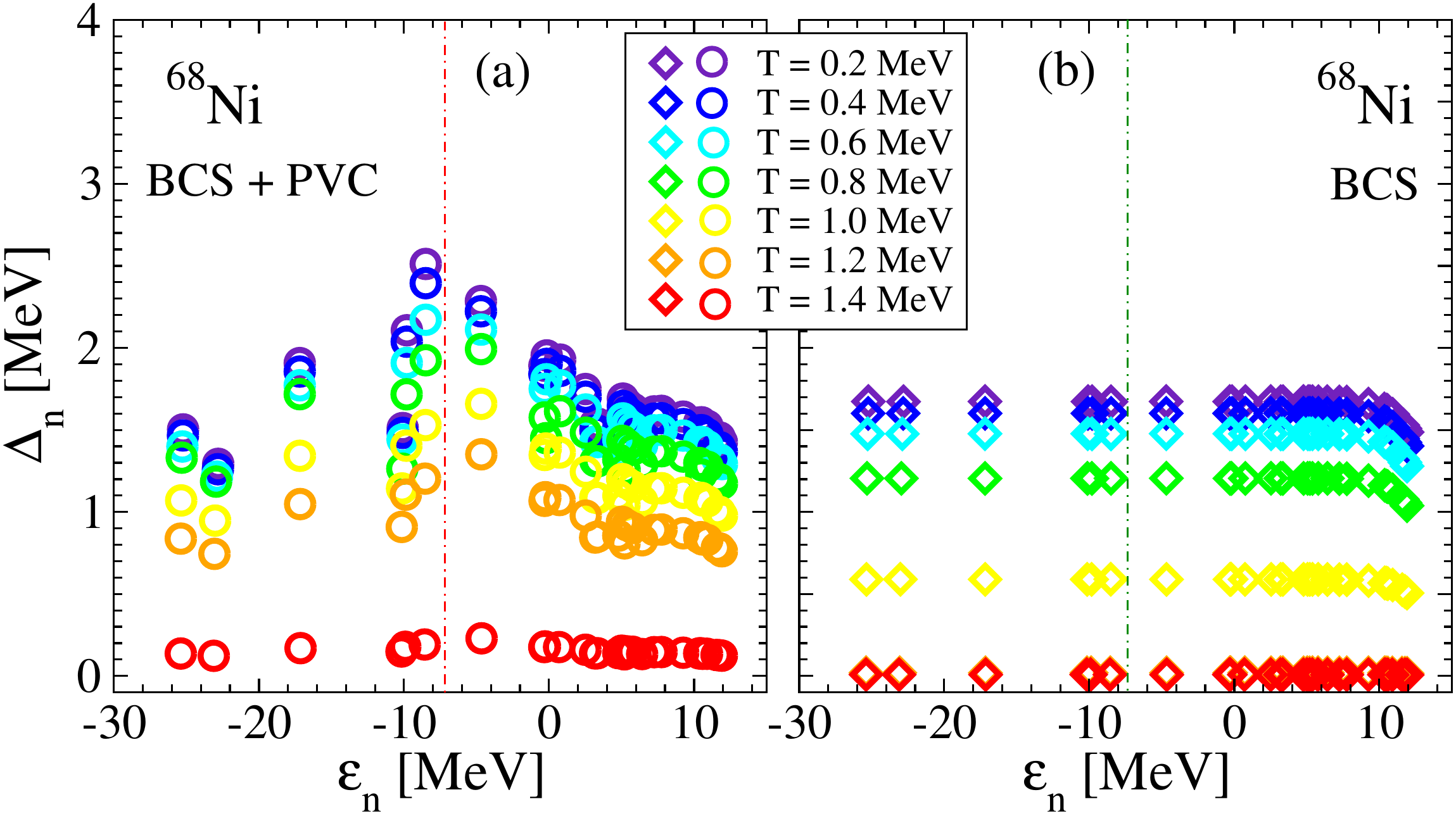}
\end{center}
\caption{Temperature dependence of the pairing gaps in $^{68}$Ni in BCS+PVC (a) compared to the pure BCS (b) approach. In both cases, the pairing gaps are gradually decreasing with the temperature increase. The vertical lines correspond to the Fermi energies in the BCS+PVC and BCS calculations at $T = 0$.}
\label{pvc_T}%
\end{figure}

\begin{figure}
\begin{center}
\includegraphics[scale=0.32]{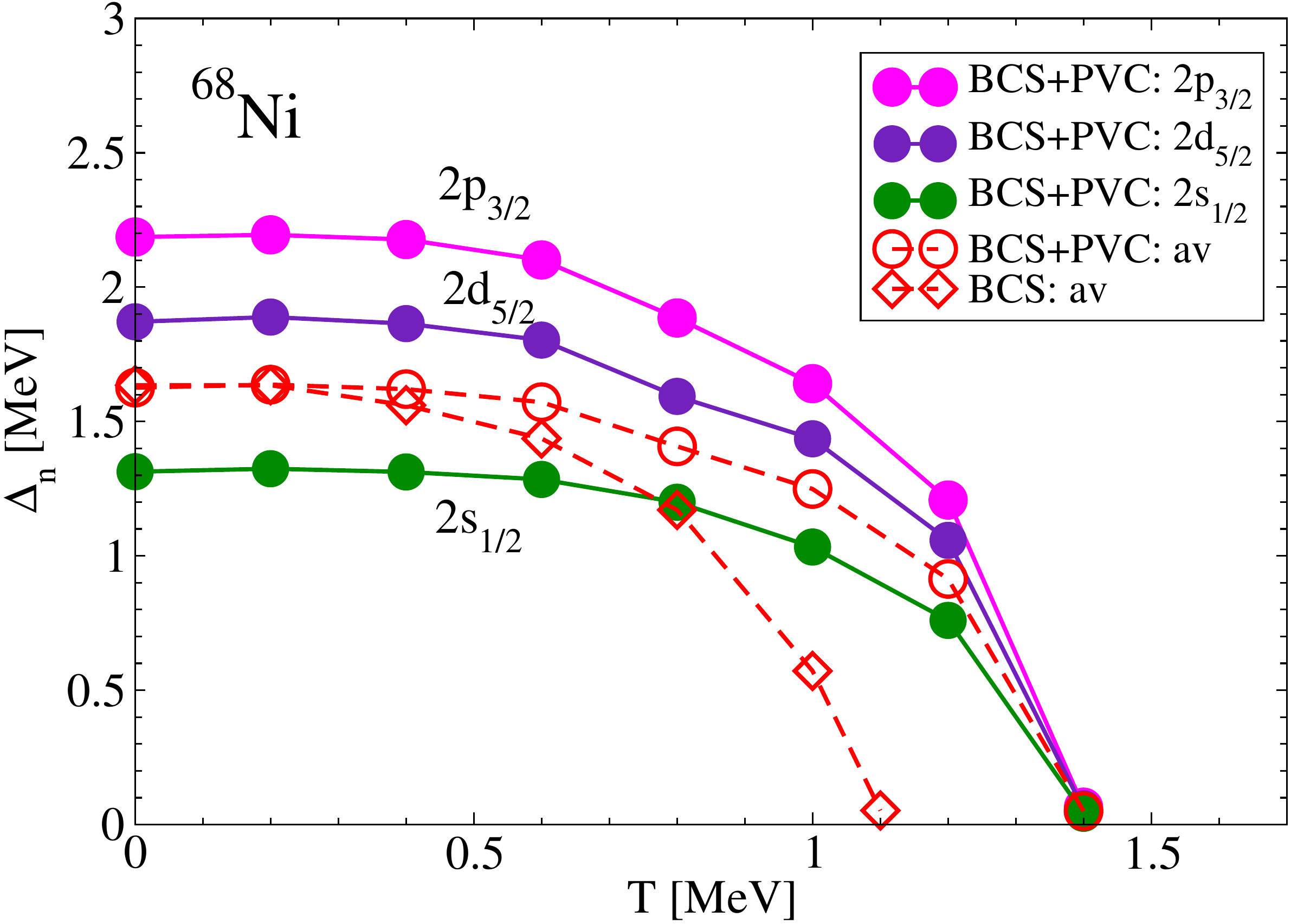}
\end{center}
\caption{Pairing gaps for the states around Fermi energy (filled symbols) and the average pairing gap (empty circles) as functions of temperature in BCS+PVC approach compared to the average pairing gap in BCS model (empty diamonds) in $^{68}$Ni.}
\label{gap_T}%
\end{figure}

The solutions obtained for the neutron pairing gap in $^{68}$Ni at low temperature of $T = 0.2$ MeV are displayed in Fig. \ref{g-sensitivity}. At this temperature the result is nearly identical to that at $T = 0$. In this calculation we illustrate (i) the role of the PVC effects in the formation of the pairing gap in the present calculation scheme and (ii) the sensitivity of the pairing gap values to the parameter $G$, which defines the strength of the static part of the interaction kernel ${\cal K}^{(0)}$.  As in the previous work of Ref. \cite{LitvinovaSchuck2020}, the latter kernel was taken in the form of the monopole-monopole interaction, which is given in detail, for instance, in Ref. \cite{LitvinovaRingTselyaev2008}, so that the present study is fully focused on the features of the dynamical kernel ${\cal K}^{(r)}$. The parameter $G$ is the only free parameter used in solving the pairing gap equation (\ref{gap}) and, ideally, will be eliminated in the ab-initio calculations, where the static kernel is determined explicitly according to Eq. (\ref{K0}). Here we adjust this parameter to reproduce on average the experimental value of the pairing gap obtained in the BCS+PVC calculation, where the averaging is weighted with the orbital degeneracy $2j+1$. The experimental value of the neutron pairing gap in $^{68}$Ni was extracted from the mass tables of Ref. \cite{Audi2002} with the aid of the commonly used three-point formula \cite{Bender2000}. 
Such calculation scheme allows for understanding the role of the dynamical PVC effects in the formation of the pairing gap. One can see from Fig. \ref{g-sensitivity} that the dynamical PVC is responsible for more than 50 \%  of the pairing gap value. Its contribution is slightly above 50 \% in the peripheral energy regions with respect to the Fermi energy (FE) and increases to 60-70\% for the states close to the FE, where the pairing gap values exhibit a smooth maximum. The presence of such maximum is attributed to the functional form of the dynamical kernel ${\cal K}^{(r)}$, namely its propagator structure. This result is qualitatively consistent with our previous work \cite{LitvinovaSchuck2020}, where a different calculations scheme was employed, as pointed out above, and with the results of Ref. \cite{BarrancoBrogliaGoriEtAl1999,Terasaki2002}. As in the case of nuclear matter \cite{Shen2005,Cao2006,Ding2016,Ramanan2018,Sedrakian2019,Guo2019a,Urban2020}, the dynamical kernel is sensitive to details of the approximation made and to the calculation scheme.
In the present implementation the experimental value of the pairing gap at $T = 0$ and low temperatures $T\leq 0.4$ MeV is best reproduced in BCS+PVC approach at $G = 12.6$ MeV, so that with this parameter value $\langle \Delta_n\rangle \approx 1.6$ MeV. The latter value of $G$ is adopted for the BCS+PVC calculations at higher temperatures, which are presented in Figs. \ref{pvc_T} and \ref{gap_T} and discussed below.

Fig. \ref{pvc_T} illustrates the temperature dependence of the neutron pairing gap in $^{68}$Ni in the BCS+PVC approach in comparison with the pure BCS model. For the latter case the parameter $G$ was increased accordingly to reproduce the experimental pairing gap, and the calculated pairing gap is almost state-independent except for the energy window border, where it smoothly decreases to the zero value because of the "soft pairing window" implied in the monopole forces \cite{LitvinovaRingTselyaev2008}. In this way, it is possible to illuminate the difference in the temperature evolution between the descriptions with only the static and with both static and dynamical kernels. In the conventional BCS, where only the static kernel is taken into account, the temperature dependence is fully determined by the factor $1-2f_2(T)$ in Eq. (\ref{gap}), while the static kernel itself has no explicit temperature dependence. Implicitly, this kernel depends on temperature, if its matrix elements are computed in a self-consistent cycle, but this dependence is relatively weak. Some additional temperature dependence may originate from the two-body density if the static kernel is calculated microscopically via Eq. (\ref{K0}). Otherwise, from the transformation (\ref{RGFtransform}) it follows that the explicit temperature dependence of the interaction kernel is the consequence of the time dependence, i.e., of the retardation effects present in the dynamical kernel components (\ref{K11}-\ref{K22}). For this study, the pairing gaps were calculated at temperatures $0 \leq T \leq 1.4$ MeV with the step of 0.2 MeV. One can see from Fig. \ref{pvc_T} that in the BCS case the pairing gap decreases quickly with the temperature increase: at $T = 1.2$ MeV it already disappears, while in the BCS+PVC calculation the gap values are still quite sizable grouping around 1 MeV. Another observation is that the pairing gap retains its peaked character in the BCS+PVC approach even when its average value decreases with the temperature increase.

The critical temperature in the theory of superfluidity is known as the temperature, at which the pairing gap vanishes. The canonical BCS relationship between the critical temperature and the value of the pairing gap at $T =0$ $\Delta_0$ is $T_c \approx 0.6\Delta_0$. In our BCS calculation, the pairing gap disappears at the temperature below $\approx 1.1$ MeV, so that the coefficient between $\Delta_0$ and $T_c$ is close to the canonical value. In the BSC+PVC approach with the additional temperature dependence of the dynamical kernel one could expect a different ratio between the $\Delta_0$ and $T_c$ values and also a variation of this ratio from state to state. These trends are illustrated in Fig. \ref{gap_T}, where we display the pairing gap as a function of temperature for selected neutron states  
in $^{68}$Ni, which was obtained in the BCS+PVC calculations. Namely, we show this function for the examples of states near and far from the Fermi energy. The average pairing gap is also shown, and these results are compared to the average BCS pairing gap. This representation helps determine more accurately the values of the critical temperatures for all the cases. The first observation from Fig. \ref{gap_T} is that in the BCS+PVC calculations the pairing gaps for all the states collapse at the same critical temperature. In the approximation described above we get the value $T_c \approx 1.4$ MeV for all the states, independently on the pairing gap values for these states at $T = 0$. Another observation is that the critical temperature in the approach with the dynamical kernel is higher than the BCS critical temperature at the same values of the $T = 0$ average pairing gaps. This indicates that the retardation effects in the dynamical component of the in-medium nucleon-nucleon interaction can help superfluidity survive at higher temperatures than it is expected in simpler models with only the static kernels. 

\begin{figure}
\begin{center}
\includegraphics[scale=0.36]{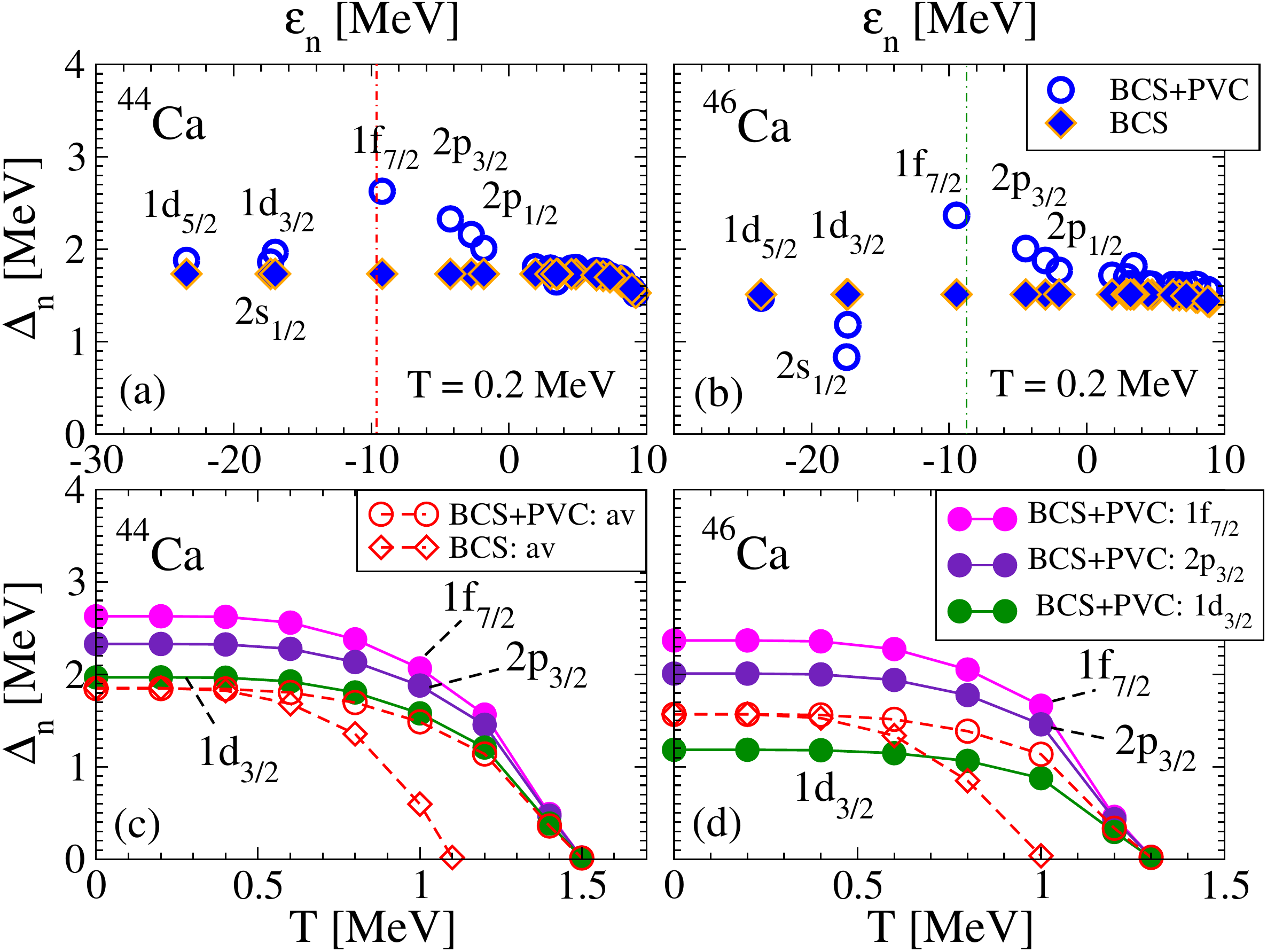}
\end{center}
\caption{Pairing gaps as functions of temperature in  $^{44,46}$Ca. Conventions are the same as in Figs. \ref{g-sensitivity} and \ref{gap_T}.}
\label{ca-gap-states_T}%
\end{figure}

\begin{figure}
\begin{center}
\includegraphics[scale=0.38]{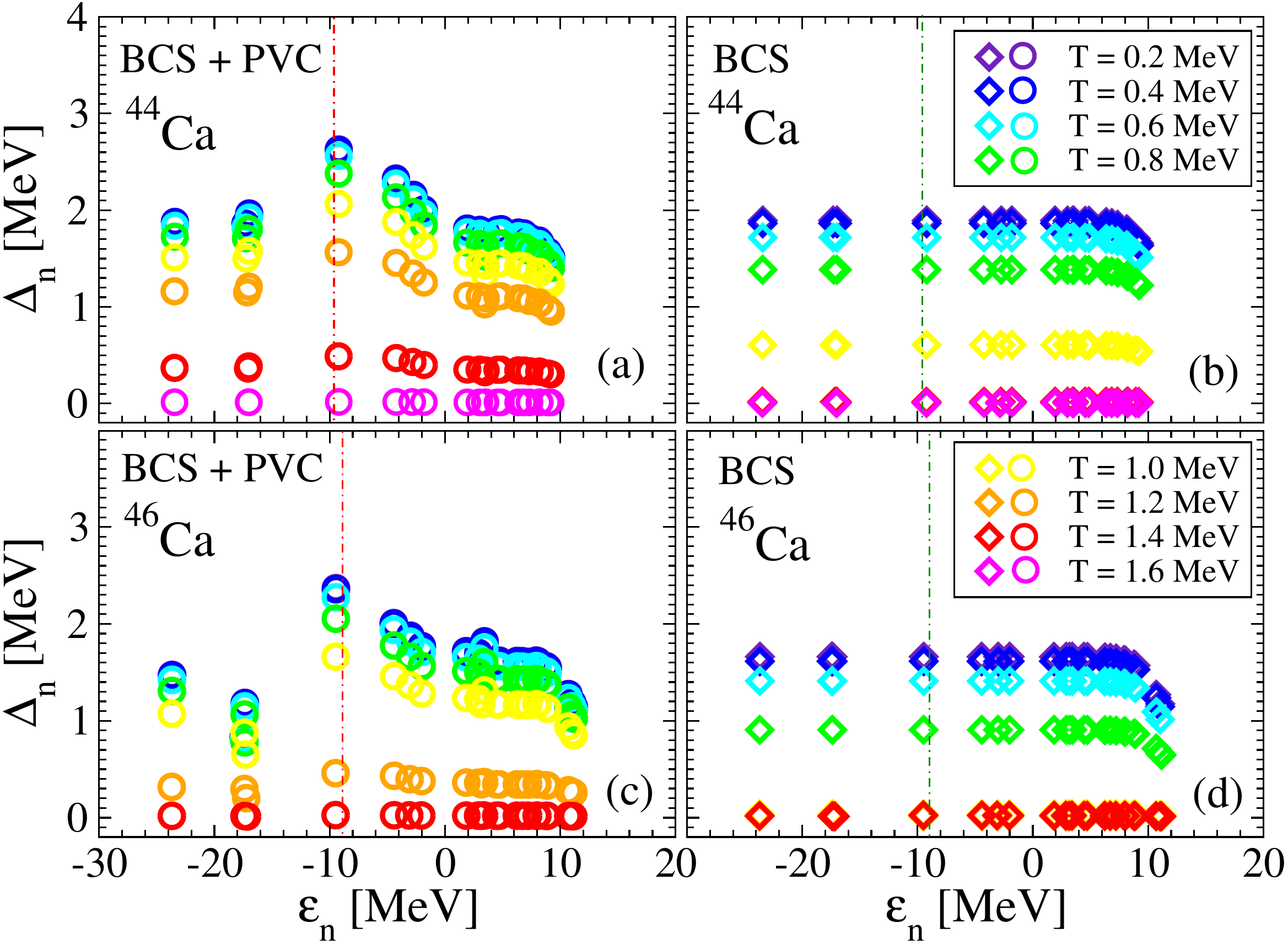}
\end{center}
\caption{Pairing gaps as functions of temperature in  $^{44,46}$Ca. Conventions are the same as in Fig. \ref{pvc_T}.}
\label{ca-gap_T}%
\end{figure}

Figs. \ref{ca-gap-states_T} and \ref{ca-gap_T} display the analogous calculations for two calcium isotopes, $^{44}$Ca and $^{46}$Ca. The upper panels of Fig.  \ref{ca-gap-states_T} show the results for the pairing gaps in these two nuclei obtained within BCS+PVC and pure BCS approximation at $T = 0.2$ MeV, which are nearly identical to $T =0$ results. Both calculations are performed with the same static pairing strength (only slightly different for $^{44}$Ca and $^{46}$Ca), in order to isolate the PVC effects. In contrast to the case of $^{68}$Ni, in calcium isotopes the PVC produces peaks in the pairing gaps mainly around the Fermi energy while affecting very little the pairing gaps of the peripheral states. As in Ref. \cite{LitvinovaSchuck2020}, this may be due to a stronger cancellation between the self-energy and exchange PVC terms for the peripheral states in these nuclei. One can also notice some irregularities in the BCS+PVC pairing gaps in $^{46}$Ca, namely the small gaps at 1d$_{3/2}$ and 2s$_{1/2}$. The remaining variance with Ref. \cite{LitvinovaSchuck2020} pertains to the differences in the calculation schemes, that is discussed below. The lower panels show the temperature dependence of the BCS+PVC pairing gaps for the states closest to the Fermi energy and their weighted averages in comparison to the behavior of the pure static pairing gaps. The latter are computed with readjusted strength to reproduce the experimental pairing gaps at $T = 0$. These calculations illustrate the drastic difference in the superfluid phase transition temperature in the two models. Similarly to the case of $^{68}$Ni, in the BCS+PVC calculations the critical temperature is considerably higher than that in the BCS, with approximately the same ratio between the BCS+PVC and BCS values. Thus, this feature of the dynamical kernel remains robust also in calcium isotopes.

In Fig. \ref{ca-gap_T} one can see a more global description of the temperature dependencies of the pairing gaps in  $^{44}$Ca and $^{46}$Ca. As in the previous case, the BCS+PVC and pure BCS calculations were performed with different values of the static pairing strength, so that the weighted average pairing gap values reproduce the $T = 0$ data in both approaches. The general result here is a noticeably slower temperature evolution of the BCS+PVC pairing gaps, as compared to the BCS ones, because of the presence of the non-trivial temperature dependence in the dynamical kernel in the former case.  This is observed for all states under study. Remarkably, the shape of the pairing gap as function of energy evolves with temperature becoming less and less peaked in the Fermi energy region with the temperature growth. 

As briefly mentioned above, the calculation scheme employed in this work is somewhat different from the one of Ref. \cite{LitvinovaSchuck2020} for $T = 0$ calculations.
There are technical reasons why we do not adopt the same calculation scheme here. In the latter scheme, on the first step, we initiated the procedure by running the RMF+BCS calculations. The BCS equation was solved self-consistently with the RMF in the usual static-kernel approximation with the monopole forces and pairing strength adjusted to the odd-even mass differences, since the phonon vertices are not yet available at this step. On the second step, the obtained pairing gap was subsequently used in the relativistic QRPA (RQRPA), and the phonon vertices and frequencies were extracted. These vertices and frequencies were used on the final step for solving the gap equation in the BCS+PVC approximation with both static and dynamical kernels. Notice that in this scheme, using the RQRPA phonons is only partially self-consistent, because in the final BCS+PVC calculation the strength constant of the static pairing kernel has to be refitted to reproduce the final pairing gap. In addition, the denominators of the dynamical kernel contain the pure single-particle energies (no quasiparticle energies) in combination with RQRPA phonon frequencies, which have the information about superfluid pairing.

To adopt an analogous calculation scheme at finite temperature, we would need to (i) generalize the RQRPA to finite temperature, which is a non-trivial task on its own, (ii) take into account a more complex structure of the phonon vertices, which means (iii) re-deriving the entire approach for the pairing propagator and, thus, for the pairing gap in the thermal RMF+BCS (or Hartree-(Fock)-Bogoliubov) quasiparticle basis. This would be essentially a more complicated and advanced solution at finite temperature, which goes beyond the scope of the present article, but which will be considered in the future. 

To avoid such complications, in this work we did not include pairing correlations on the first two steps, and, instead, ran thermal RMF and finite-temperature relativistic RPA to obtain the mean-field and phonon characteristics. On one hand, this reduces the accuracy of the phonon calculations at T = 0, but, on the other hand, allows us to avoid the inconsistency between the denominators of the dynamical kernel, which would contain the energies and frequencies obtained in the quasiparticle picture, and exponential factors with the temperature dependence, which are not yet adopted to superfluid pairing. 

These differences in the calculation schemes, which are both approximate and may be replaced by a more accurate one in the future, are responsible for somewhat different behavior of the resulting pairing gaps obtained in this work at T = 0, as compared to Ref. \cite{LitvinovaSchuck2020}.  However, their enhancements for the states surrounding the Fermi energy and the general trends in nickel and calcium nuclei remain similar, although in the Ref. \cite{LitvinovaSchuck2020} the peaks of the pairing gaps are not always centered at the Fermi surface. This may be because of only partial self-consistency described above, which modifies the location of the poles of the dynamical kernel. Slightly weaker PVC effects in Ref. \cite{LitvinovaSchuck2020} are observed mainly due to the reduction of the PVC vertices, as they are multiplied with the combinations of the occupation factors in RQRPA \cite{LitvinovaRingTselyaev2008}.

\vspace{0.3cm}
\section{Summary and outlook}
\label{Conclusions}

The equation of motion for the two-time two-fermion correlation function in a strongly coupled many-body system at finite temperature is considered. We show that, as in the case of zero temperature, 
the EOM for this propagator takes the form of the Dyson-Bethe-Salpeter equation with the interaction kernel, which is split into the static and dynamical components. This kernel includes, in principle, all the in-medium physics derived from the underlying bare two-fermion interaction. While the static component of the kernel depends on the correlated two-fermionic density, the dynamical component contains a higher-rank fermionic propagator. The latter, in the case of the symmetric form of the kernel, is represented by the propagator of four fermions. Factorization of this propagator allows for the truncation of the many-body problem at the level of two-body correlation functions whose EOM's, together with those for the one-fermion correlation function discussed in Ref. \cite{LitvinovaSchuck2019} form a closed system of integral equations. The equation for the temperature-dependent pairing gap, which is related to the residue of the two-time particle-particle propagator, is formulated as a static limit of the EOM for this propagator. To extract the pairing gap, which is one of the most common characteristics of superfluidity, we adopt the BCS-like variational ansatz for the ground state wave function. The resulting equation allows for an extension of the BCS approximation to correlations of higher complexity, which introduce an additional non-trivial temperature dependence of the pairing gap.

The effects of the dynamical kernel at finite temperature are illustrated in the calculations of the neutron pairing gaps for $^{68}$Ni and two calcium isotopes, $^{44}$Ca and $^{46}$Ca. The pole character of this kernel gives rise to the peak of the pairing gap around the Fermi surface at all temperatures when the pairing gap has non-zero values. We find that the time dependence, mostly the retardation, present in this kernel translates to a different temperature dependence of the pairing gap than the one of the BCS approximation. In particular, the presence of the dynamical term leads to noticeably higher values of the critical temperature. This finding may be important for numerous applications. For instance, in applications to nuclear astrophysics, such as the r-process nucleosynthesis in the neutron star mergers and the supernovae explosion, the nuclear input for temperatures below 1-2 MeV is involved. Crossing the critical temperature, i.e., the superfluid phase transition can affect considerably the excitation spectra snd, thus, various reaction rates. An example of the electron capture is discussed in Ref. \cite{LitvinovaRobin2021}. Therefore, the form of the dynamical kernel and its temperature dependence has to be computed as accurate as possible. Self-consistent calculations of this kernel as well as the static one, desirably in an ab-initio framework, thus, remain an important topic for future research.


\section*{Acknowledgements}
This work is supported by the US-NSF Career Grant PHY-1654379.
%

\bibliography{Bibliography_Dec2020}

\begin{thebibliography}{51}%
\makeatletter
\providecommand \@ifxundefined [1]{%
 \@ifx{#1\undefined}
}%
\providecommand \@ifnum [1]{%
 \ifnum #1\expandafter \@firstoftwo
 \else \expandafter \@secondoftwo
 \fi
}%
\providecommand \@ifx [1]{%
 \ifx #1\expandafter \@firstoftwo
 \else \expandafter \@secondoftwo
 \fi
}%
\providecommand \natexlab [1]{#1}%
\providecommand \enquote  [1]{``#1''}%
\providecommand \bibnamefont  [1]{#1}%
\providecommand \bibfnamefont [1]{#1}%
\providecommand \citenamefont [1]{#1}%
\providecommand \href@noop [0]{\@secondoftwo}%
\providecommand \href [0]{\begingroup \@sanitize@url \@href}%
\providecommand \@href[1]{\@@startlink{#1}\@@href}%
\providecommand \@@href[1]{\endgroup#1\@@endlink}%
\providecommand \@sanitize@url [0]{\catcode `\\12\catcode `\$12\catcode
  `\&12\catcode `\#12\catcode `\^12\catcode `\_12\catcode `\%12\relax}%
\providecommand \@@startlink[1]{}%
\providecommand \@@endlink[0]{}%
\providecommand \url  [0]{\begingroup\@sanitize@url \@url }%
\providecommand \@url [1]{\endgroup\@href {#1}{\urlprefix }}%
\providecommand \urlprefix  [0]{URL }%
\providecommand \Eprint [0]{\href }%
\providecommand \doibase [0]{http://dx.doi.org/}%
\providecommand \selectlanguage [0]{\@gobble}%
\providecommand \bibinfo  [0]{\@secondoftwo}%
\providecommand \bibfield  [0]{\@secondoftwo}%
\providecommand \translation [1]{[#1]}%
\providecommand \BibitemOpen [0]{}%
\providecommand \bibitemStop [0]{}%
\providecommand \bibitemNoStop [0]{.\EOS\space}%
\providecommand \EOS [0]{\spacefactor3000\relax}%
\providecommand \BibitemShut  [1]{\csname bibitem#1\endcsname}%
\let\auto@bib@innerbib\@empty
\bibitem [{\citenamefont {Bohr}\ \emph {et~al.}(1958)\citenamefont {Bohr},
  \citenamefont {Mottelson},\ and\ \citenamefont {Pines}}]{Bohr1958}%
  \BibitemOpen
  \bibfield  {author} {\bibinfo {author} {\bibfnamefont {A.}~\bibnamefont
  {Bohr}}, \bibinfo {author} {\bibfnamefont {B.~R.}\ \bibnamefont {Mottelson}},
  \ and\ \bibinfo {author} {\bibfnamefont {D.}~\bibnamefont {Pines}},\ }\href
  {\doibase 10.1103/PhysRev.110.936} {\bibfield  {journal} {\bibinfo  {journal}
  {Physical Review}\ }\textbf {\bibinfo {volume} {110}},\ \bibinfo {pages}
  {936} (\bibinfo {year} {1958})}\BibitemShut {NoStop}%
\bibitem [{\citenamefont {Bardeen}\ \emph {et~al.}(1957)\citenamefont
  {Bardeen}, \citenamefont {Cooper},\ and\ \citenamefont
  {Schrieffer}}]{Bardeen1957}%
  \BibitemOpen
  \bibfield  {author} {\bibinfo {author} {\bibfnamefont {J.}~\bibnamefont
  {Bardeen}}, \bibinfo {author} {\bibfnamefont {L.~N.}\ \bibnamefont {Cooper}},
  \ and\ \bibinfo {author} {\bibfnamefont {J.~R.}\ \bibnamefont {Schrieffer}},\
  }\href {\doibase 10.1103/PhysRev.106.162} {\bibfield  {journal} {\bibinfo
  {journal} {Physical Review}\ }\textbf {\bibinfo {volume} {106}},\ \bibinfo
  {pages} {162} (\bibinfo {year} {1957})}\BibitemShut {NoStop}%
\bibitem [{\citenamefont {Broglia}\ and\ \citenamefont
  {Zelevinsky}(2013)}]{50BCS}%
  \BibitemOpen
  \bibinfo {editor} {\bibfnamefont {R.}~\bibnamefont {Broglia}}\ and\ \bibinfo
  {editor} {\bibfnamefont {V.}~\bibnamefont {Zelevinsky}},\ eds.,\ \href@noop
  {} {\emph {\bibinfo {title} {Fifty Years Of Nuclear BCS: Pairing In Finite
  Systems}}}\ (\bibinfo  {publisher} {World Scientific},\ \bibinfo {year}
  {2013})\BibitemShut {NoStop}%
\bibitem [{\citenamefont {Shen}\ \emph {et~al.}(2005)\citenamefont {Shen},
  \citenamefont {Lombardo},\ and\ \citenamefont {Schuck}}]{Shen2005}%
  \BibitemOpen
  \bibfield  {author} {\bibinfo {author} {\bibfnamefont {C.}~\bibnamefont
  {Shen}}, \bibinfo {author} {\bibfnamefont {U.}~\bibnamefont {Lombardo}}, \
  and\ \bibinfo {author} {\bibfnamefont {P.}~\bibnamefont {Schuck}},\ }\href
  {\doibase 10.1103/PhysRevC.71.054301} {\bibfield  {journal} {\bibinfo
  {journal} {Physical Review C}\ }\textbf {\bibinfo {volume} {71}},\ \bibinfo
  {pages} {054301} (\bibinfo {year} {2005})}\BibitemShut {NoStop}%
\bibitem [{\citenamefont {Cao}\ \emph {et~al.}(2006)\citenamefont {Cao},
  \citenamefont {Lombardo},\ and\ \citenamefont {Schuck}}]{Cao2006}%
  \BibitemOpen
  \bibfield  {author} {\bibinfo {author} {\bibfnamefont {L.~G.}\ \bibnamefont
  {Cao}}, \bibinfo {author} {\bibfnamefont {U.}~\bibnamefont {Lombardo}}, \
  and\ \bibinfo {author} {\bibfnamefont {P.}~\bibnamefont {Schuck}},\ }\href
  {\doibase 10.1103/PhysRevC.74.064301} {\bibfield  {journal} {\bibinfo
  {journal} {Physical Review C}\ }\textbf {\bibinfo {volume} {74}},\ \bibinfo
  {pages} {064301} (\bibinfo {year} {2006})}\BibitemShut {NoStop}%
\bibitem [{\citenamefont {Ding}\ \emph {et~al.}(2016)\citenamefont {Ding},
  \citenamefont {Rios}, \citenamefont {Dussan}, \citenamefont {Dickhoff},
  \citenamefont {Witte}, \citenamefont {Carbone},\ and\ \citenamefont
  {Polls}}]{Ding2016}%
  \BibitemOpen
  \bibfield  {author} {\bibinfo {author} {\bibfnamefont {D.}~\bibnamefont
  {Ding}}, \bibinfo {author} {\bibfnamefont {A.}~\bibnamefont {Rios}}, \bibinfo
  {author} {\bibfnamefont {H.}~\bibnamefont {Dussan}}, \bibinfo {author}
  {\bibfnamefont {W.~H.}\ \bibnamefont {Dickhoff}}, \bibinfo {author}
  {\bibfnamefont {S.~J.}\ \bibnamefont {Witte}}, \bibinfo {author}
  {\bibfnamefont {A.}~\bibnamefont {Carbone}}, \ and\ \bibinfo {author}
  {\bibfnamefont {A.}~\bibnamefont {Polls}},\ }\href {\doibase
  10.1103/PhysRevC.94.025802} {\bibfield  {journal} {\bibinfo  {journal}
  {Physical Review C}\ }\textbf {\bibinfo {volume} {94}},\ \bibinfo {pages}
  {025802} (\bibinfo {year} {2016})}\BibitemShut {NoStop}%
\bibitem [{\citenamefont {Ramanan}\ and\ \citenamefont
  {Urban}(2018)}]{Ramanan2018}%
  \BibitemOpen
  \bibfield  {author} {\bibinfo {author} {\bibfnamefont {S.}~\bibnamefont
  {Ramanan}}\ and\ \bibinfo {author} {\bibfnamefont {M.}~\bibnamefont
  {Urban}},\ }\href {\doibase 10.1103/PhysRevC.98.024314} {\bibfield  {journal}
  {\bibinfo  {journal} {Physical Review C}\ }\textbf {\bibinfo {volume} {98}},\
  \bibinfo {pages} {024314} (\bibinfo {year} {2018})}\BibitemShut {NoStop}%
\bibitem [{\citenamefont {Sedrakian}\ and\ \citenamefont
  {Clark}(2019)}]{Sedrakian2019}%
  \BibitemOpen
  \bibfield  {author} {\bibinfo {author} {\bibfnamefont {A.}~\bibnamefont
  {Sedrakian}}\ and\ \bibinfo {author} {\bibfnamefont {J.}~\bibnamefont
  {Clark}},\ }\href@noop {} {\bibfield  {journal} {\bibinfo  {journal}
  {European Physical Journal}\ }\textbf {\bibinfo {volume} {A55}},\ \bibinfo
  {pages} {167} (\bibinfo {year} {2019})}\BibitemShut {NoStop}%
\bibitem [{\citenamefont {Guo}\ \emph {et~al.}(2019)\citenamefont {Guo},
  \citenamefont {Lombardo},\ and\ \citenamefont {Schuck}}]{Guo2019a}%
  \BibitemOpen
  \bibfield  {author} {\bibinfo {author} {\bibfnamefont {W.}~\bibnamefont
  {Guo}}, \bibinfo {author} {\bibfnamefont {U.}~\bibnamefont {Lombardo}}, \
  and\ \bibinfo {author} {\bibfnamefont {P.}~\bibnamefont {Schuck}},\ }\href
  {\doibase 10.1103/PhysRevC.99.014310} {\bibfield  {journal} {\bibinfo
  {journal} {Physical Review}\ }\textbf {\bibinfo {volume} {99}},\ \bibinfo
  {pages} {014310} (\bibinfo {year} {2019})}\BibitemShut {NoStop}%
\bibitem [{\citenamefont {Urban}\ and\ \citenamefont
  {Ramanan}(2020)}]{Urban2020}%
  \BibitemOpen
  \bibfield  {author} {\bibinfo {author} {\bibfnamefont {M.}~\bibnamefont
  {Urban}}\ and\ \bibinfo {author} {\bibfnamefont {S.}~\bibnamefont
  {Ramanan}},\ }\href@noop {} {\bibfield  {journal} {\bibinfo  {journal}
  {Physical Review C}\ }\textbf {\bibinfo {volume} {101}},\ \bibinfo {pages}
  {035803} (\bibinfo {year} {2020})}\BibitemShut {NoStop}%
\bibitem [{\citenamefont {Bohr}\ and\ \citenamefont
  {Mottelson}(1969)}]{BohrMottelson1969}%
  \BibitemOpen
  \bibfield  {author} {\bibinfo {author} {\bibfnamefont {A.}~\bibnamefont
  {Bohr}}\ and\ \bibinfo {author} {\bibfnamefont {B.~R.}\ \bibnamefont
  {Mottelson}},\ }\href@noop {} {\emph {\bibinfo {title} {Nuclear
  structure}}},\ Vol.~\bibinfo {volume} {1}\ (\bibinfo  {publisher} {World
  Scientific},\ \bibinfo {year} {1969})\BibitemShut {NoStop}%
\bibitem [{\citenamefont {Bohr}\ and\ \citenamefont
  {Mottelson}(1975)}]{BohrMottelson1975}%
  \BibitemOpen
  \bibfield  {author} {\bibinfo {author} {\bibfnamefont {A.}~\bibnamefont
  {Bohr}}\ and\ \bibinfo {author} {\bibfnamefont {B.~R.}\ \bibnamefont
  {Mottelson}},\ }\href@noop {} {\emph {\bibinfo {title} {Nuclear
  structure}}},\ Vol.~\bibinfo {volume} {2}\ (\bibinfo  {publisher} {Benjamin,
  New York},\ \bibinfo {year} {1975})\BibitemShut {NoStop}%
\bibitem [{\citenamefont {Van~der Sluys}\ \emph {et~al.}(1993)\citenamefont
  {Van~der Sluys}, \citenamefont {Van~Neck}, \citenamefont {Waroquier},\ and\
  \citenamefont {Ryckebusch}}]{VanderSluys1993}%
  \BibitemOpen
  \bibfield  {author} {\bibinfo {author} {\bibfnamefont {V.}~\bibnamefont
  {Van~der Sluys}}, \bibinfo {author} {\bibfnamefont {D.}~\bibnamefont
  {Van~Neck}}, \bibinfo {author} {\bibfnamefont {M.}~\bibnamefont {Waroquier}},
  \ and\ \bibinfo {author} {\bibfnamefont {J.}~\bibnamefont {Ryckebusch}},\
  }\href {\doibase 10.1016/0375-9474(93)90479-H} {\bibfield  {journal}
  {\bibinfo  {journal} {Nuclear Physics}\ }\textbf {\bibinfo {volume} {A551}},\
  \bibinfo {pages} {210} (\bibinfo {year} {1993})}\BibitemShut {NoStop}%
\bibitem [{\citenamefont {Avdeenkov}\ and\ \citenamefont
  {Kamerdzhiev}(1999)}]{Avdeenkov1999}%
  \BibitemOpen
  \bibfield  {author} {\bibinfo {author} {\bibfnamefont {A.~V.}\ \bibnamefont
  {Avdeenkov}}\ and\ \bibinfo {author} {\bibfnamefont {S.~P.}\ \bibnamefont
  {Kamerdzhiev}},\ }\href {\doibase 10.1016/S0370-2693(99)00719-4} {\bibfield
  {journal} {\bibinfo  {journal} {Physics Letters}\ }\textbf {\bibinfo {volume}
  {B459}},\ \bibinfo {pages} {423} (\bibinfo {year} {1999})}\BibitemShut
  {NoStop}%
\bibitem [{\citenamefont {Avdeenkov}\ and\ \citenamefont
  {Kamerdzhev}(1999)}]{Avdeenkov1999a}%
  \BibitemOpen
  \bibfield  {author} {\bibinfo {author} {\bibfnamefont {A.~V.}\ \bibnamefont
  {Avdeenkov}}\ and\ \bibinfo {author} {\bibfnamefont {S.~P.}\ \bibnamefont
  {Kamerdzhev}},\ }\href {\doibase 10.1134/1.568080} {\bibfield  {journal}
  {\bibinfo  {journal} {JETP Letters}\ }\textbf {\bibinfo {volume} {69}},\
  \bibinfo {pages} {715} (\bibinfo {year} {1999})}\BibitemShut {NoStop}%
\bibitem [{\citenamefont {Barranco}\ \emph {et~al.}(1999)\citenamefont
  {Barranco}, \citenamefont {Broglia}, \citenamefont {Gori}, \citenamefont
  {Vigezzi}, \citenamefont {Bortignon},\ and\ \citenamefont
  {Terasaki}}]{BarrancoBrogliaGoriEtAl1999}%
  \BibitemOpen
  \bibfield  {author} {\bibinfo {author} {\bibfnamefont {F.}~\bibnamefont
  {Barranco}}, \bibinfo {author} {\bibfnamefont {R.}~\bibnamefont {Broglia}},
  \bibinfo {author} {\bibfnamefont {G.}~\bibnamefont {Gori}}, \bibinfo {author}
  {\bibfnamefont {E.}~\bibnamefont {Vigezzi}}, \bibinfo {author} {\bibfnamefont
  {P.}~\bibnamefont {Bortignon}}, \ and\ \bibinfo {author} {\bibfnamefont
  {J.}~\bibnamefont {Terasaki}},\ }\href@noop {} {\bibfield  {journal}
  {\bibinfo  {journal} {Physical Review Letters}\ }\textbf {\bibinfo {volume}
  {83}},\ \bibinfo {pages} {2147} (\bibinfo {year} {1999})}\BibitemShut
  {NoStop}%
\bibitem [{\citenamefont {Barranco}\ \emph {et~al.}(2005)\citenamefont
  {Barranco}, \citenamefont {Bortignon}, \citenamefont {Broglia}, \citenamefont
  {Col{\`o}}, \citenamefont {Schuck}, \citenamefont {Vigezzi},\ and\
  \citenamefont {Vinas}}]{BarrancoBortignonBrogliaEtAl2005}%
  \BibitemOpen
  \bibfield  {author} {\bibinfo {author} {\bibfnamefont {F.}~\bibnamefont
  {Barranco}}, \bibinfo {author} {\bibfnamefont {P.}~\bibnamefont {Bortignon}},
  \bibinfo {author} {\bibfnamefont {R.}~\bibnamefont {Broglia}}, \bibinfo
  {author} {\bibfnamefont {G.}~\bibnamefont {Col{\`o}}}, \bibinfo {author}
  {\bibfnamefont {P.}~\bibnamefont {Schuck}}, \bibinfo {author} {\bibfnamefont
  {E.}~\bibnamefont {Vigezzi}}, \ and\ \bibinfo {author} {\bibfnamefont
  {X.}~\bibnamefont {Vinas}},\ }\href@noop {} {\bibfield  {journal} {\bibinfo
  {journal} {Physical Review C}\ }\textbf {\bibinfo {volume} {72}},\ \bibinfo
  {pages} {054314} (\bibinfo {year} {2005})}\BibitemShut {NoStop}%
\bibitem [{\citenamefont {Idini}\ \emph {et~al.}(2015)\citenamefont {Idini},
  \citenamefont {Potel}, \citenamefont {Barranco}, \citenamefont {Vigezzi},\
  and\ \citenamefont {Broglia}}]{IdiniPotelBarrancoEtAl2015}%
  \BibitemOpen
  \bibfield  {author} {\bibinfo {author} {\bibfnamefont {A.}~\bibnamefont
  {Idini}}, \bibinfo {author} {\bibfnamefont {G.}~\bibnamefont {Potel}},
  \bibinfo {author} {\bibfnamefont {F.}~\bibnamefont {Barranco}}, \bibinfo
  {author} {\bibfnamefont {E.}~\bibnamefont {Vigezzi}}, \ and\ \bibinfo
  {author} {\bibfnamefont {R.~A.}\ \bibnamefont {Broglia}},\ }\href@noop {}
  {\bibfield  {journal} {\bibinfo  {journal} {Physical Review C}\ }\textbf
  {\bibinfo {volume} {92}},\ \bibinfo {pages} {031304} (\bibinfo {year}
  {2015})}\BibitemShut {NoStop}%
\bibitem [{\citenamefont {Cooper}(2010)}]{50BCSorig}%
  \BibitemOpen
  \bibinfo {editor} {\bibfnamefont {L.~N.}\ \bibnamefont {Cooper}},\ ed.,\
  \href@noop {} {\emph {\bibinfo {title} {BCS: 50 Years}}}\ (\bibinfo
  {publisher} {World Scientific},\ \bibinfo {year} {2010})\BibitemShut
  {NoStop}%
\bibitem [{\citenamefont {Strinati}\ \emph {et~al.}(2018)\citenamefont
  {Strinati}, \citenamefont {Pieri}, \citenamefont {R\"opke}, \citenamefont
  {Schuck},\ and\ \citenamefont {Urban}}]{Strinati2018}%
  \BibitemOpen
  \bibfield  {author} {\bibinfo {author} {\bibfnamefont {G.~C.}\ \bibnamefont
  {Strinati}}, \bibinfo {author} {\bibfnamefont {P.}~\bibnamefont {Pieri}},
  \bibinfo {author} {\bibfnamefont {G.}~\bibnamefont {R\"opke}}, \bibinfo
  {author} {\bibfnamefont {P.}~\bibnamefont {Schuck}}, \ and\ \bibinfo {author}
  {\bibfnamefont {M.}~\bibnamefont {Urban}},\ }\href {\doibase
  https://doi.org/10.1016/j.physrep.2018.02.004} {\bibfield  {journal}
  {\bibinfo  {journal} {Physics Reports}\ }\textbf {\bibinfo {volume} {738}},\
  \bibinfo {pages} {1 } (\bibinfo {year} {2018})}\BibitemShut {NoStop}%
\bibitem [{\citenamefont {Langanke}\ \emph {et~al.}(1996)\citenamefont
  {Langanke}, \citenamefont {Dean}, \citenamefont {Radha},\ and\ \citenamefont
  {Koonin}}]{Langanke1996}%
  \BibitemOpen
  \bibfield  {author} {\bibinfo {author} {\bibfnamefont {K.}~\bibnamefont
  {Langanke}}, \bibinfo {author} {\bibfnamefont {D.}~\bibnamefont {Dean}},
  \bibinfo {author} {\bibfnamefont {P.}~\bibnamefont {Radha}}, \ and\ \bibinfo
  {author} {\bibfnamefont {S.}~\bibnamefont {Koonin}},\ }\href {\doibase
  https://doi.org/10.1016/0375-9474(96)00139-X} {\bibfield  {journal} {\bibinfo
   {journal} {Nuclear Physics}\ }\textbf {\bibinfo {volume} {A602}},\ \bibinfo
  {pages} {244 } (\bibinfo {year} {1996})}\BibitemShut {NoStop}%
\bibitem [{\citenamefont {Liu}\ and\ \citenamefont
  {Alhassid}(2001)}]{LiuAlhassid2001}%
  \BibitemOpen
  \bibfield  {author} {\bibinfo {author} {\bibfnamefont {S.}~\bibnamefont
  {Liu}}\ and\ \bibinfo {author} {\bibfnamefont {Y.}~\bibnamefont {Alhassid}},\
  }\href {\doibase 10.1103/PhysRevLett.87.022501} {\bibfield  {journal}
  {\bibinfo  {journal} {Physical Review Letters}\ }\textbf {\bibinfo {volume}
  {87}},\ \bibinfo {pages} {022501} (\bibinfo {year} {2001})}\BibitemShut
  {NoStop}%
\bibitem [{\citenamefont {Jin}\ \emph {et~al.}(2010)\citenamefont {Jin},
  \citenamefont {Urban},\ and\ \citenamefont {Schuck}}]{Jin2010}%
  \BibitemOpen
  \bibfield  {author} {\bibinfo {author} {\bibfnamefont {M.}~\bibnamefont
  {Jin}}, \bibinfo {author} {\bibfnamefont {M.}~\bibnamefont {Urban}}, \ and\
  \bibinfo {author} {\bibfnamefont {P.}~\bibnamefont {Schuck}},\ }\href
  {\doibase 10.1103/PhysRevC.82.024911} {\bibfield  {journal} {\bibinfo
  {journal} {Physical Review C}\ }\textbf {\bibinfo {volume} {82}},\ \bibinfo
  {pages} {024911} (\bibinfo {year} {2010})}\BibitemShut {NoStop}%
\bibitem [{\citenamefont {Dukelsky}\ \emph {et~al.}(1998)\citenamefont
  {Dukelsky}, \citenamefont {R\"opke},\ and\ \citenamefont
  {Schuck}}]{DukelskyRoepkeSchuck1998}%
  \BibitemOpen
  \bibfield  {author} {\bibinfo {author} {\bibfnamefont {J.}~\bibnamefont
  {Dukelsky}}, \bibinfo {author} {\bibfnamefont {G.}~\bibnamefont {R\"opke}}, \
  and\ \bibinfo {author} {\bibfnamefont {P.}~\bibnamefont {Schuck}},\ }\href
  {\doibase 10.1016/S0375-9474(97)00606-4} {\bibfield  {journal} {\bibinfo
  {journal} {Nuclear Physics}\ }\textbf {\bibinfo {volume} {A628}},\ \bibinfo
  {pages} {17} (\bibinfo {year} {1998})}\BibitemShut {NoStop}%
\bibitem [{\citenamefont {Storozhenko}\ \emph {et~al.}(2003)\citenamefont
  {Storozhenko}, \citenamefont {Schuck}, \citenamefont {Dukelsky},
  \citenamefont {R{\"o}pke},\ and\ \citenamefont {Vdovin}}]{Storozhenko2003}%
  \BibitemOpen
  \bibfield  {author} {\bibinfo {author} {\bibfnamefont {A.}~\bibnamefont
  {Storozhenko}}, \bibinfo {author} {\bibfnamefont {P.}~\bibnamefont {Schuck}},
  \bibinfo {author} {\bibfnamefont {J.}~\bibnamefont {Dukelsky}}, \bibinfo
  {author} {\bibfnamefont {G.}~\bibnamefont {R{\"o}pke}}, \ and\ \bibinfo
  {author} {\bibfnamefont {A.}~\bibnamefont {Vdovin}},\ }\href {\doibase
  10.1016/S0003-4916(03)00095-2} {\bibfield  {journal} {\bibinfo  {journal}
  {Annals of Physics}\ }\textbf {\bibinfo {volume} {307}},\ \bibinfo {pages}
  {308} (\bibinfo {year} {2003})}\BibitemShut {NoStop}%
\bibitem [{\citenamefont {Tiago}\ \emph {et~al.}(2008)\citenamefont {Tiago},
  \citenamefont {Kent}, \citenamefont {Hood},\ and\ \citenamefont
  {Reboredo}}]{Tiago2008}%
  \BibitemOpen
  \bibfield  {author} {\bibinfo {author} {\bibfnamefont {M.~L.}\ \bibnamefont
  {Tiago}}, \bibinfo {author} {\bibfnamefont {P.}~\bibnamefont {Kent}},
  \bibinfo {author} {\bibfnamefont {R.~Q.}\ \bibnamefont {Hood}}, \ and\
  \bibinfo {author} {\bibfnamefont {F.~A.}\ \bibnamefont {Reboredo}},\
  }\href@noop {} {\bibfield  {journal} {\bibinfo  {journal} {Journal of
  Chemical Physics}\ }\textbf {\bibinfo {volume} {129}},\ \bibinfo {pages}
  {084311} (\bibinfo {year} {2008})}\BibitemShut {NoStop}%
\bibitem [{\citenamefont {Martinez}\ \emph {et~al.}(2010)\citenamefont
  {Martinez}, \citenamefont {Garc{\'\i}a-Lastra}, \citenamefont {L{\'o}pez},\
  and\ \citenamefont {Alonso}}]{Martinez2010}%
  \BibitemOpen
  \bibfield  {author} {\bibinfo {author} {\bibfnamefont {J.~I.}\ \bibnamefont
  {Martinez}}, \bibinfo {author} {\bibfnamefont {J.}~\bibnamefont
  {Garc{\'\i}a-Lastra}}, \bibinfo {author} {\bibfnamefont {M.}~\bibnamefont
  {L{\'o}pez}}, \ and\ \bibinfo {author} {\bibfnamefont {J.}~\bibnamefont
  {Alonso}},\ }\href@noop {} {\bibfield  {journal} {\bibinfo  {journal}
  {Journal of Chemical Physics}\ }\textbf {\bibinfo {volume} {132}},\ \bibinfo
  {pages} {044314} (\bibinfo {year} {2010})}\BibitemShut {NoStop}%
\bibitem [{\citenamefont {Sangalli}\ \emph {et~al.}(2011)\citenamefont
  {Sangalli}, \citenamefont {Romaniello}, \citenamefont {Onida},\ and\
  \citenamefont {Marini}}]{Sangalli2011}%
  \BibitemOpen
  \bibfield  {author} {\bibinfo {author} {\bibfnamefont {D.}~\bibnamefont
  {Sangalli}}, \bibinfo {author} {\bibfnamefont {P.}~\bibnamefont
  {Romaniello}}, \bibinfo {author} {\bibfnamefont {G.}~\bibnamefont {Onida}}, \
  and\ \bibinfo {author} {\bibfnamefont {A.}~\bibnamefont {Marini}},\
  }\href@noop {} {\bibfield  {journal} {\bibinfo  {journal} {Journal of
  Chemical Physics}\ }\textbf {\bibinfo {volume} {134}},\ \bibinfo {pages}
  {034115} (\bibinfo {year} {2011})}\BibitemShut {NoStop}%
\bibitem [{\citenamefont {Schuck}\ and\ \citenamefont
  {Tohyama}(2016)}]{SchuckTohyama2016}%
  \BibitemOpen
  \bibfield  {author} {\bibinfo {author} {\bibfnamefont {P.}~\bibnamefont
  {Schuck}}\ and\ \bibinfo {author} {\bibfnamefont {M.}~\bibnamefont
  {Tohyama}},\ }\href {\doibase 10.1140/epja/i2016-16307-7} {\bibfield
  {journal} {\bibinfo  {journal} {European Physical Journal A}\ }\textbf
  {\bibinfo {volume} {52}},\ \bibinfo {pages} {307} (\bibinfo {year}
  {2016})}\BibitemShut {NoStop}%
\bibitem [{\citenamefont {Olevano}\ \emph {et~al.}(2018)\citenamefont
  {Olevano}, \citenamefont {Toulouse},\ and\ \citenamefont
  {Schuck}}]{Olevano2018}%
  \BibitemOpen
  \bibfield  {author} {\bibinfo {author} {\bibfnamefont {V.}~\bibnamefont
  {Olevano}}, \bibinfo {author} {\bibfnamefont {J.}~\bibnamefont {Toulouse}}, \
  and\ \bibinfo {author} {\bibfnamefont {P.}~\bibnamefont {Schuck}},\ }\href
  {\doibase 10.1063/1.5080330} {\bibfield  {journal} {\bibinfo  {journal}
  {Journal of Chemical Physics}\ }\textbf {\bibinfo {volume} {150}},\ \bibinfo
  {pages} {084112} (\bibinfo {year} {2018})}\BibitemShut {NoStop}%
\bibitem [{\citenamefont {Litvinova}\ and\ \citenamefont
  {Schuck}(2020)}]{LitvinovaSchuck2020}%
  \BibitemOpen
  \bibfield  {author} {\bibinfo {author} {\bibfnamefont {E.}~\bibnamefont
  {Litvinova}}\ and\ \bibinfo {author} {\bibfnamefont {P.}~\bibnamefont
  {Schuck}},\ }\href {\doibase 10.1103/PhysRevC.102.034310} {\bibfield
  {journal} {\bibinfo  {journal} {Physical Review C}\ }\textbf {\bibinfo
  {volume} {102}},\ \bibinfo {pages} {034310} (\bibinfo {year}
  {2020})}\BibitemShut {NoStop}%
\bibitem [{\citenamefont {Matsubara}(1955)}]{Matsubara1955}%
  \BibitemOpen
  \bibfield  {author} {\bibinfo {author} {\bibfnamefont {T.}~\bibnamefont
  {Matsubara}},\ }\href@noop {} {\bibfield  {journal} {\bibinfo  {journal}
  {Progress in Theoretical Physics}\ }\textbf {\bibinfo {volume} {14}},\
  \bibinfo {pages} {351} (\bibinfo {year} {1955})}\BibitemShut {NoStop}%
\bibitem [{\citenamefont {Zagoskin}(2014)}]{Zagoskin2014}%
  \BibitemOpen
  \bibfield  {author} {\bibinfo {author} {\bibfnamefont {A.}~\bibnamefont
  {Zagoskin}},\ }\href@noop {} {\emph {\bibinfo {title} {Quantum Theory of
  Many-Body Systems, Graduate Texts in Physics}}}\ (\bibinfo  {publisher}
  {Springer International Publishing, Cham},\ \bibinfo {year}
  {2014})\BibitemShut {NoStop}%
\bibitem [{\citenamefont {Abrikosov}\ \emph {et~al.}(1975)\citenamefont
  {Abrikosov}, \citenamefont {Gorkov},\ and\ \citenamefont
  {Dzyaloshinski}}]{Abrikosov1975}%
  \BibitemOpen
  \bibfield  {author} {\bibinfo {author} {\bibfnamefont {A.~A.}\ \bibnamefont
  {Abrikosov}}, \bibinfo {author} {\bibfnamefont {L.~P.}\ \bibnamefont
  {Gorkov}}, \ and\ \bibinfo {author} {\bibfnamefont {I.~E.}\ \bibnamefont
  {Dzyaloshinski}},\ }\href@noop {} {\emph {\bibinfo {title} {Methods of
  Quantum FieldTheory in Statistical Physics}}}\ (\bibinfo  {publisher} {Dover
  Publications, New York},\ \bibinfo {year} {1975})\BibitemShut {NoStop}%
\bibitem [{\citenamefont {Litvinova}\ and\ \citenamefont
  {Schuck}(2019)}]{LitvinovaSchuck2019}%
  \BibitemOpen
  \bibfield  {author} {\bibinfo {author} {\bibfnamefont {E.}~\bibnamefont
  {Litvinova}}\ and\ \bibinfo {author} {\bibfnamefont {P.}~\bibnamefont
  {Schuck}},\ }\href@noop {} {\bibfield  {journal} {\bibinfo  {journal}
  {Physical Review C}\ }\textbf {\bibinfo {volume} {100}},\ \bibinfo {pages}
  {064320} (\bibinfo {year} {2019})}\BibitemShut {NoStop}%
\bibitem [{\citenamefont {Schuck}(2019)}]{Schuck2019}%
  \BibitemOpen
  \bibfield  {author} {\bibinfo {author} {\bibfnamefont {P.}~\bibnamefont
  {Schuck}},\ }\href {\doibase 10.1140/epja/i2019-12798-x} {\bibfield
  {journal} {\bibinfo  {journal} {European Physical Journal}\ }\textbf
  {\bibinfo {volume} {A55}},\ \bibinfo {pages} {250} (\bibinfo {year}
  {2019})}\BibitemShut {NoStop}%
\bibitem [{\citenamefont {Ring}\ and\ \citenamefont
  {Schuck}(1980)}]{RingSchuck1980}%
  \BibitemOpen
  \bibfield  {author} {\bibinfo {author} {\bibfnamefont {P.}~\bibnamefont
  {Ring}}\ and\ \bibinfo {author} {\bibfnamefont {P.}~\bibnamefont {Schuck}},\
  }\href@noop {} {\emph {\bibinfo {title} {The Nuclear Many-Body Problem}}}\
  (\bibinfo  {publisher} {Springer-Verlag Berlin Heidelberg},\ \bibinfo {year}
  {1980})\BibitemShut {NoStop}%
\bibitem [{\citenamefont {Rabhi}\ \emph {et~al.}(2002)\citenamefont {Rabhi},
  \citenamefont {Bennaceur}, \citenamefont {Chanfray},\ and\ \citenamefont
  {Schuck}}]{Rabhi2002}%
  \BibitemOpen
  \bibfield  {author} {\bibinfo {author} {\bibfnamefont {A.}~\bibnamefont
  {Rabhi}}, \bibinfo {author} {\bibfnamefont {R.}~\bibnamefont {Bennaceur}},
  \bibinfo {author} {\bibfnamefont {G.}~\bibnamefont {Chanfray}}, \ and\
  \bibinfo {author} {\bibfnamefont {P.}~\bibnamefont {Schuck}},\ }\href
  {\doibase 10.1103/PhysRevC.66.064315} {\bibfield  {journal} {\bibinfo
  {journal} {Physical Review C}\ }\textbf {\bibinfo {volume} {66}},\ \bibinfo
  {pages} {064315} (\bibinfo {year} {2002})}\BibitemShut {NoStop}%
\bibitem [{\citenamefont {Robin}\ and\ \citenamefont
  {Litvinova}(2016)}]{RobinLitvinova2016}%
  \BibitemOpen
  \bibfield  {author} {\bibinfo {author} {\bibfnamefont {C.}~\bibnamefont
  {Robin}}\ and\ \bibinfo {author} {\bibfnamefont {E.}~\bibnamefont
  {Litvinova}},\ }\href@noop {} {\bibfield  {journal} {\bibinfo  {journal}
  {European Physical Journal A}\ }\textbf {\bibinfo {volume} {52}},\ \bibinfo
  {pages} {205} (\bibinfo {year} {2016})}\BibitemShut {NoStop}%
\bibitem [{\citenamefont {Robin}\ and\ \citenamefont
  {Litvinova}(2019)}]{Robin2019}%
  \BibitemOpen
  \bibfield  {author} {\bibinfo {author} {\bibfnamefont {C.}~\bibnamefont
  {Robin}}\ and\ \bibinfo {author} {\bibfnamefont {E.}~\bibnamefont
  {Litvinova}},\ }\href@noop {} {\bibfield  {journal} {\bibinfo  {journal}
  {Physical Review Letters}\ }\textbf {\bibinfo {volume} {123}},\ \bibinfo
  {pages} {202501} (\bibinfo {year} {2019})}\BibitemShut {NoStop}%
\bibitem [{\citenamefont {Ainsworth}\ \emph {et~al.}(1989)\citenamefont
  {Ainsworth}, \citenamefont {Wambach},\ and\ \citenamefont
  {Pines}}]{Ainsworth1989}%
  \BibitemOpen
  \bibfield  {author} {\bibinfo {author} {\bibfnamefont {T.}~\bibnamefont
  {Ainsworth}}, \bibinfo {author} {\bibfnamefont {J.}~\bibnamefont {Wambach}},
  \ and\ \bibinfo {author} {\bibfnamefont {D.}~\bibnamefont {Pines}},\ }\href
  {\doibase https://doi.org/10.1016/0370-2693(89)91246-X} {\bibfield  {journal}
  {\bibinfo  {journal} {Physics Letters B}\ }\textbf {\bibinfo {volume}
  {222}},\ \bibinfo {pages} {173 } (\bibinfo {year} {1989})}\BibitemShut
  {NoStop}%
\bibitem [{\citenamefont {Schulze}\ \emph {et~al.}(1996)\citenamefont
  {Schulze}, \citenamefont {Cugnon}, \citenamefont {Lejeune}, \citenamefont
  {Baldo},\ and\ \citenamefont {Lombardo}}]{Schulze1996}%
  \BibitemOpen
  \bibfield  {author} {\bibinfo {author} {\bibfnamefont {H.-J.}\ \bibnamefont
  {Schulze}}, \bibinfo {author} {\bibfnamefont {J.}~\bibnamefont {Cugnon}},
  \bibinfo {author} {\bibfnamefont {A.}~\bibnamefont {Lejeune}}, \bibinfo
  {author} {\bibfnamefont {M.}~\bibnamefont {Baldo}}, \ and\ \bibinfo {author}
  {\bibfnamefont {U.}~\bibnamefont {Lombardo}},\ }\href {\doibase
  https://doi.org/10.1016/0370-2693(96)00213-4} {\bibfield  {journal} {\bibinfo
   {journal} {Physics Letters B}\ }\textbf {\bibinfo {volume} {375}},\ \bibinfo
  {pages} {1 } (\bibinfo {year} {1996})}\BibitemShut {NoStop}%
\bibitem [{\citenamefont {Lalazissis}\ \emph {et~al.}(1997)\citenamefont
  {Lalazissis}, \citenamefont {K{\"o}nig},\ and\ \citenamefont
  {Ring}}]{Lalazissis1997}%
  \BibitemOpen
  \bibfield  {author} {\bibinfo {author} {\bibfnamefont {G.}~\bibnamefont
  {Lalazissis}}, \bibinfo {author} {\bibfnamefont {J.}~\bibnamefont
  {K{\"o}nig}}, \ and\ \bibinfo {author} {\bibfnamefont {P.}~\bibnamefont
  {Ring}},\ }\href@noop {} {\bibfield  {journal} {\bibinfo  {journal} {Physical
  Review C}\ }\textbf {\bibinfo {volume} {55}},\ \bibinfo {pages} {540}
  (\bibinfo {year} {1997})}\BibitemShut {NoStop}%
\bibitem [{\citenamefont {Litvinova}\ and\ \citenamefont
  {Wibowo}(2018)}]{LitvinovaWibowo2018}%
  \BibitemOpen
  \bibfield  {author} {\bibinfo {author} {\bibfnamefont {E.}~\bibnamefont
  {Litvinova}}\ and\ \bibinfo {author} {\bibfnamefont {H.}~\bibnamefont
  {Wibowo}},\ }\href@noop {} {\bibfield  {journal} {\bibinfo  {journal}
  {Physical Review Letters}\ }\textbf {\bibinfo {volume} {121}},\ \bibinfo
  {pages} {082501} (\bibinfo {year} {2018})}\BibitemShut {NoStop}%
\bibitem [{\citenamefont {Wibowo}\ and\ \citenamefont
  {Litvinova}(2019)}]{WibowoLitvinova2019}%
  \BibitemOpen
  \bibfield  {author} {\bibinfo {author} {\bibfnamefont {H.}~\bibnamefont
  {Wibowo}}\ and\ \bibinfo {author} {\bibfnamefont {E.}~\bibnamefont
  {Litvinova}},\ }\href@noop {} {\bibfield  {journal} {\bibinfo  {journal}
  {Physical Review C}\ }\textbf {\bibinfo {volume} {100}},\ \bibinfo {pages}
  {024307} (\bibinfo {year} {2019})}\BibitemShut {NoStop}%
\bibitem [{\citenamefont {Litvinova}\ and\ \citenamefont
  {Wibowo}(2019)}]{LitvinovaWibowo2019}%
  \BibitemOpen
  \bibfield  {author} {\bibinfo {author} {\bibfnamefont {E.}~\bibnamefont
  {Litvinova}}\ and\ \bibinfo {author} {\bibfnamefont {H.}~\bibnamefont
  {Wibowo}},\ }\href@noop {} {\bibfield  {journal} {\bibinfo  {journal}
  {European Physical Journal}\ }\textbf {\bibinfo {volume} {A55}},\ \bibinfo
  {pages} {223} (\bibinfo {year} {2019})}\BibitemShut {NoStop}%
\bibitem [{\citenamefont {Litvinova}\ \emph {et~al.}(2008)\citenamefont
  {Litvinova}, \citenamefont {Ring},\ and\ \citenamefont
  {Tselyaev}}]{LitvinovaRingTselyaev2008}%
  \BibitemOpen
  \bibfield  {author} {\bibinfo {author} {\bibfnamefont {E.}~\bibnamefont
  {Litvinova}}, \bibinfo {author} {\bibfnamefont {P.}~\bibnamefont {Ring}}, \
  and\ \bibinfo {author} {\bibfnamefont {V.}~\bibnamefont {Tselyaev}},\
  }\href@noop {} {\bibfield  {journal} {\bibinfo  {journal} {Physical Review
  C}\ }\textbf {\bibinfo {volume} {78}},\ \bibinfo {pages} {014312} (\bibinfo
  {year} {2008})}\BibitemShut {NoStop}%
\bibitem [{\citenamefont {Audi}\ \emph {et~al.}(2002)\citenamefont {Audi},
  \citenamefont {Wapstra},\ and\ \citenamefont {Thibault}}]{Audi2002}%
  \BibitemOpen
  \bibfield  {author} {\bibinfo {author} {\bibfnamefont {G.}~\bibnamefont
  {Audi}}, \bibinfo {author} {\bibfnamefont {A.~H.}\ \bibnamefont {Wapstra}}, \
  and\ \bibinfo {author} {\bibfnamefont {C.}~\bibnamefont {Thibault}},\ }\href
  {\doibase 10.1016/j.nuclphysa.2003.11.003} {\bibfield  {journal} {\bibinfo
  {journal} {Nuclear Physics}\ }\textbf {\bibinfo {volume} {A729}},\ \bibinfo
  {pages} {337} (\bibinfo {year} {2002})}\BibitemShut {NoStop}%
\bibitem [{\citenamefont {Bender}\ \emph {et~al.}(2000)\citenamefont {Bender},
  \citenamefont {Rutz}, \citenamefont {Reinhard},\ and\ \citenamefont
  {Maruhn}}]{Bender2000}%
  \BibitemOpen
  \bibfield  {author} {\bibinfo {author} {\bibfnamefont {M.}~\bibnamefont
  {Bender}}, \bibinfo {author} {\bibfnamefont {K.}~\bibnamefont {Rutz}},
  \bibinfo {author} {\bibfnamefont {P.~G.}\ \bibnamefont {Reinhard}}, \ and\
  \bibinfo {author} {\bibfnamefont {J.~A.}\ \bibnamefont {Maruhn}},\ }\href
  {\doibase 10.1007/s10050-000-4504-z} {\bibfield  {journal} {\bibinfo
  {journal} {European Physical Journal}\ }\textbf {\bibinfo {volume} {A8}},\
  \bibinfo {pages} {59} (\bibinfo {year} {2000})}\BibitemShut {NoStop}%
\bibitem [{\citenamefont {Terasaki}\ \emph {et~al.}(2002)\citenamefont
  {Terasaki}, \citenamefont {Barranco}, \citenamefont {Bortignon},
  \citenamefont {Broglia},\ and\ \citenamefont {Vigezzi}}]{Terasaki2002}%
  \BibitemOpen
  \bibfield  {author} {\bibinfo {author} {\bibfnamefont {J.}~\bibnamefont
  {Terasaki}}, \bibinfo {author} {\bibfnamefont {F.}~\bibnamefont {Barranco}},
  \bibinfo {author} {\bibfnamefont {P.~F.}\ \bibnamefont {Bortignon}}, \bibinfo
  {author} {\bibfnamefont {R.~A.}\ \bibnamefont {Broglia}}, \ and\ \bibinfo
  {author} {\bibfnamefont {E.}~\bibnamefont {Vigezzi}},\ }\href {\doibase
  10.1143/PTP.108.495} {\bibfield  {journal} {\bibinfo  {journal} {Progress in
  Theoretical Physics}\ }\textbf {\bibinfo {volume} {108}},\ \bibinfo {pages}
  {495} (\bibinfo {year} {2002})}\BibitemShut {NoStop}%
\bibitem [{\citenamefont {Litvinova}\ and\ \citenamefont
  {Robin}(2021)}]{LitvinovaRobin2021}%
  \BibitemOpen
  \bibfield  {author} {\bibinfo {author} {\bibfnamefont {E.}~\bibnamefont
  {Litvinova}}\ and\ \bibinfo {author} {\bibfnamefont {C.}~\bibnamefont
  {Robin}},\ }\href {\doibase 10.1103/PhysRevC.103.024326} {\bibfield
  {journal} {\bibinfo  {journal} {Physical Review C}\ }\textbf {\bibinfo
  {volume} {103}},\ \bibinfo {pages} {024326} (\bibinfo {year}
  {2021})}\BibitemShut {NoStop}%
\end{thebibliography}%
\end{document}